\documentclass[a4paper,11pt]{article}
\pdfoutput=1 

\usepackage{graphicx} 
\usepackage{subcaption}
\usepackage{array}
\newcolumntype{P}[1]{>{\centering\arraybackslash}p{#1}}
\usepackage{floatrow}
\graphicspath{ {./images/} }

\usepackage{jinstpub} 

\usepackage{amsmath}  
\usepackage{lineno}

\usepackage[utf8]{inputenc}
\usepackage{siunitx}

\title{Wavelength-shifter coated polystyrene as an easy-to-build and low-cost plastic scintillator detector}
\author[a]{A. Brignoli,}
\author[a]{A. Conaboy,}
\author[c]{V. Dormenev,}
\author[b,1]{D. Jimeno,\note{Corresponding author.}}
\author[c]{D. Kazlou,}
\author[a]{H. Lacker,}
\author[a]{C. Scharf,}
\author[a]{J. Schmidt,}
\author[c]{H. G. Zaunick}

\emailAdd{doramasjs.17@gmail.com}

\affiliation[a]{Institut f\"ur Physik, Humboldt-Universit\"at zu Berlin, 
\\Newtonstr. 15, 12489 Berlin, Germany}
\affiliation[b]{Facultat de Física , Universitat de Barcelona, 
\\Martí i Franquès 1, 08028 Barcelona, Spain}
\affiliation[c]{II. Physikalisches Institut, Justus-Liebig-Universität, 
\\Heinrich-Buff-Ring 16, 35392 Giessen, Germany}

\date{October 2022}

\abstract{We studied the light yield of a pure polystyrene slide coated with wavelength-shifter molecules, coupled to a photomultiplier, using $\beta^{-}$ particles from a $^{90}$Sr source, as a possible easy-to-build, low-cost plastic scintillator detector. Comparison measurements were performed with an uncoated polystyrene slide as well as with uncoated and coated PMMA slides, the latter which can only produce Cherenkov light when being traversed by charged particles. The results with the single (double) coated polystyrene slides show about 4.9 (6.3) times higher detected photon yield compared to the uncoated slide. For comparison, the light yield of a polystyrene-based extruded plastic scintillator material doped with PTP and POPOP was measured as well.
The absolute detected light yield motivates future studies for developing easy-to-build, low-cost polystyrene-based plastic scintillator detectors.}

\begin{document}

\maketitle

\section{Introduction}
\label{Sec:Introduction}
For several years, wavelength-shifting optical modules (WOMs) are being considered as possible photon detectors with large area and low noise and costs~\cite{Hebecker:2016mrq,Bastian-Querner:2021uqv}. They have been originally proposed as water-Cherenkov photon detectors in the context of the \hbox{Upgrades} of the IceCube neutrino telescope~\cite{Hebecker:2016mrq}. In fact, WOMs will be deployed for a seven-string extension of the IceCube detector planned for the 2022/2023 South Pole deployment season~\cite{IceCube:2021mdb}. Following the first proposal ~\cite{Hebecker:2016mrq}, they have been also proposed as photon detectors for the liquid-scintillator based Surround Background Tagger of the SHiP experiment proposal at the CERN SPS~\cite{SHiP:2015vad, SHIP:2021tpn}, with a proof-of-principle for such a WOM-based liquid-scintillator detector reported in~\cite{Ehlert:2018pke}.

A standard WOM is a large-area transparent tube, typically made of  \hbox{PMMA} or of quartz glass, coated from outside\footnote{Coating from inside can be performed as well, which increases the photon absorption probability.} with
a wavelength-shifting (WLS) dye, e.g. using a dip-coating technique~\cite{Hebecker:2016mrq,Bastian-Querner:2021uqv}. Typically, wavelength-shifting in this context refers to absorption of ultraviolet (UV) photons by the WLS dye followed by re-emission of photons in the blue range, although the principle can of course be applied as well to absorption and re-emission in other wavelength ranges.
The absorption process inside the WLS layer
has a high probability, if the WLS layer thickness is sufficiently large. If the WOM tube is placed inside a medium of much smaller refraction index, such as air, most of the secondary photons, which are emitted isotropically from the point of absorption, fulfil the condition of total reflection when hitting the inner or outer WOM-tube walls. In the absence of absorption and scattering losses, \SI{74.6}{\percent} of the secondary photons are guided to the two ends of the WOM tube \cite{Hebecker:2016mrq}.
As a result, the majority of the secondary photons can be guided
to one of the two WOM tube ends where they can be eventually detected
by a photomultiplier (PMT) that only has to cover the WOM-tube diameter or by a
ring-array of SiPMs that only has to cover the WOM tube exit area.
For the WOMs considered for the IceCube extension and the SHiP surround background
tagger, the following WLS dye, here referred as the WLS standard dye, is used: \SI{77.31}{\percent} toluene, \SI{22.29}{\percent} Paraloid B723, and as the two wavelength shifters, \SI{0.13}{\percent} bis-MSB and \SI{0.27}{\percent}  p-terphenyl (PTP) \cite{Hebecker:2016mrq}.  The absorption spectrum of the WLS dye ranges from about 250 nm up to about 400 nm and its emission spectrum ranges from about 400 nm up to 500 nm \cite{IceCube:2021mdb}. For the study presented in this paper, an additional modified dye was used. To produce the modified dye, \SI{22}{\percent} of the toluene in the standard dye was evaporated. This leads to an increase in the viscosity of the dye, which in turn increases the thickness of a dip-coated dye layer~\cite{Bastian-Querner:2021uqv}.

In this work, we modify the WOM concept in two ways: 1) Instead of a tube geometry, we consider a planar geometry, e.g. a slide. The WOM principle concerning the condition of total reflection will be preserved in a planar geometry and therefore it is expected
that many photons can be guided to the thin-sized side walls of a slide. 
2) The WOM base material serves now as the stand-alone active detector material. Hence UV photons produced by Cherenkov radiation and, depending on the material, possibly also by scintillation inside this material are wavelength-shifted and guided to the photon detector.
The idea to use a WLS layer on a transparent material for Cherenkov light detection was already considered in Ref.\cite{GrandeMoss:1983}. For our planar WOM material, we consider here in particular pure polystyrene (PS) that can be commercially purchased, at least for specific thicknesses varying between 1 and \SI{5}{\milli\meter}, at quite low cost. Pure polystyrene is not only producing Cherenkov light but is also scintillating, although with low light output, increasing the overall light yield. Pure PS has an emission spectrum from scintillation in the UV range between about 290 nm and 400 nm with an emission maximum of about \SI{320}{\nano\meter} \cite{Bross:1992pe}, which fits well to the WLS absorption spectrum. The position of the emission maximum and the fluorescence yield of polystyrene can vary between about 310 and 340 nm, depending on the concentration of the styrene monomer inside polystyrene, which depends on the polymerisation temperature, and whether it is casted or non-casted polystyrene, as well as on the degree of polymerisation cross-linking induced by an additional agent \cite{Basile:1961}.
As a result, by dip-coating a PS slide with the standard WLS dye, it is expected that a part of the UV scintillation photons can be absorbed inside the WLS dye after a relatively short travel distance, controlled by the thickness of the slide and the attenuation length of the PS material, and re-emitted as visible
photons that can be detected by a photosensor (e.g. a PMT or a SiPM). 
One can not expect light yields as large as for standard PS-based plastic scintillators doped with a primary fluorophore and with WLS molecules being mixed into the PS. However, the advantage of the material and procedure studied in our work is that one can produce without large efforts planar PS detectors by simply dip-coating a commercially available, very cheap material that can be cut into the desired slide geometries. The limitations of commercially available PS material are geometries and thicknesses. 
To demonstrate the feasibility of the concept, we studied with a $\beta^-$ source various square-shaped PS slides of \SI{50}{\milli\meter} side length and about \SI{5}{\milli\meter} thickness, being the maximum thickness found on the commercial market, and, as a comparison, also slides made of non-UV-transparent PMMA, which does not scintillate and acts only as a Cherenkov light radiator when being traversed by a high-energy charged particle. 

For comparison, we studied as well an extruded scintillator material based on polystyrene with a PTP concentration of 1.5\% and a concentration of POPOP of 0.01\% produced by the UNIPLAST plant (Vladimir, Russia) \cite{Kudenko:2001qj} that was e. g. studied as a possible base material for 3 m long scintillator bars equipped with WLS-fibres and silicon photomultipliers to be used in the muon detector \cite{Baldini:2016bxh} of the proposed SHiP experiment \cite{SHIP:2021tpn}. Such scintillators can be produced at moderate costs and generate high light-yield, however with only moderate attenuation lengths of about 10 to 15 cm in the visible range \cite{Baldini:2016bxh}, which are even smaller than the attenuation length of pure polystyrene.

\section{Experimental setup and measurements}
\label{Sec:SetupSketch}

\subsection{Sample preparation}
\label{Sec:SamplePreparation}
We studied the following samples of plastic material slides: 
UV-non-transparent PMMA without WLS coating (PMMA), 
UV-non-transparent PMMA dip-coated with the standard WLS dye (PMMA-WLS),
UV-non-transparent PMMA dip-coated twice with the modified WLS dye (PMMA-WLS-2) to enhance the WLS layer thickness, 
polystyrene without WLS coating (PS), 
polystyrene dip-coated with the standard WLS dye (PS-WLS), 
and polystyrene dip-coated twice with the modified WLS dye (PS-WLS-2).

All PS slides were cut from a larger, commercially available, $1 \times 0.5$ ${\rm m}^2$ plate of about \SI{5}{\milli\meter} thickness \footnote{The PS plate was purchased for a price of 29,45 Euro.}, and polished after cutting. 
The WLS absorption spectrum covers the range \SIrange{250}{385}{\nano\meter}, the emission spectrum starts at \SI{400}{\nano\meter} with the maximum at about \SI{420}{\nano\meter} going up to about \SI{500}{\nano\meter}. The slides were dipped into the WLS dye for 80\,s, withdrawn from the WLS dye at a speed of \SI{93}{\milli\meter\per\second} using the dip-coater ND-DC Dip Coater from the company Nadetech Innovations S.L. also used in our production of dip-coated WOM tubes, and then dried for at least \SI{24}{\hour}.

The dip-coating procedure
was optimized in dedicated studies to maximize the absorption probability of PMMA WOM tubes developed for the SHiP Surround Background Tagger \cite{JakobPaulSchmidt:MasterThesis}.
With respect to the standard WLS dye, 
\SI{22}{\percent} of the toluene from the WLS-2 dye was partially evaporated before dip-coating to enhance the dye viscosity, since this helps in producing a larger WLS layer thickness \cite{Bastian-Querner:2021uqv}. Coating with the standard paint results in WLS layers of
about only \SI{5}{\micro\meter} thickness,
as measured on dip-coated glass slides using the stylus profiler DektakXT from Bruker Corporation.
Significantly higher thicknesses can be achieved using the modified paint with the  maximal coating speed of 150 mm/min achieved with
the dip-coater in use (ND-DC Dip Coater from Nadetech Innovations S.L.) and coating the slide twice. For double-coated slides, it is important to avoid a significant immersion time of the slide in the
dye during second coating, as the toluene
dissolves the already 
existing WLS layer. With this procedure, WLS layer thicknesses on glass slides between
\SI{25}{\micro\meter} and \SI{40}{\micro\meter} have been achieved with the modified WLS dye.
Using dip-coated glass slides, also the attenuation length of the WLS layer has been 
estimated from  transmission measurements,
ranging from \SI{7}{\micro\meter} at \SI{320}{\nano\meter}  down to \SI{4}{\micro\meter} at \SI{350}{\nano\meter} and up again to \SI{8}{\micro\meter} at \SI{380}{\nano\meter}.

Determining the WLS layer thickness on a plastic material slide, either by comparing the weight of the slide before and after dip-coating or by using the profiler method applied for coated glass slides, does not work since the toluene dissolves a part of the plastic material surface during the coating process. 
Comparisons of dip-coated glass slides with extruded and casted PMMA slides however show that
the transmission behaviour with respect to
the WLS dye layer agree with each other. Therefore, similar WLS layer thicknesses can be achieved on plastic material as on glass. Since the WLS thicknesses achieved when coating twice with the modified WLS dye are much larger than the attenuation lengths, one expects a very high absorption probability of UV photons in the range between 300 and 400 nm.

To compare the performance with a high-light-yield plastic scintillator,
a slide with the same dimensions was cut out from the extruded plastic scintillator material produced by UNIPLAST.

\subsection{Sample characterization}
\label{Sec:SampleCharacterization}
In this sub-section, we describe pre-measurements to characterize the materials under study: 1. the transmission spectra through the uncoated PS slide, the uncoated PMMA slide, and the UNIPLAST reference slide were measured, from which the corresponding light attenuation lengths were determined; 2. the primary scintillation light yield in pure PS and UNIPLAST samples were estimated.

\subsubsection{Transmission measurements}
\label{Sec:TransmissionMeasurements}

\begin{figure}[!htbp]
    \centering
    \includegraphics[width=0.8\textwidth]{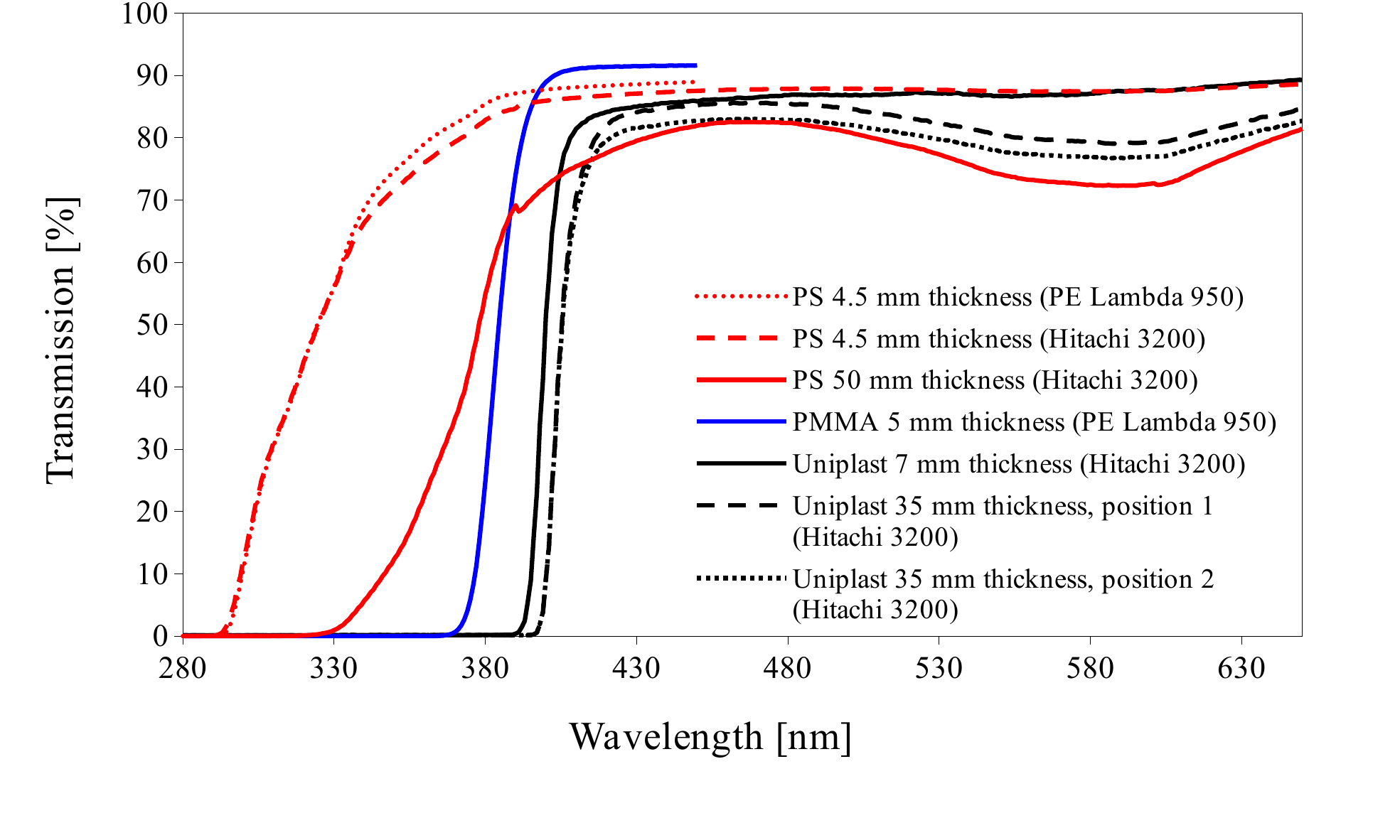}
    \caption{Transmission spectra measured with a PerkinElmer Lambda 950 spectrophotometer (PE Lambda 950) and with a Hitachi 3200 spectrophotometer (Hitachi 3200). In red: transmission spectra for the uncoated pure PS slide with light path lengths through the material of \SI{4.5}{\milli\meter} (dotted and dashed line) and \SI{50}{\milli\meter}, respectively (solid line). In blue: transmission spectrum for the uncoated PMMA slide with a light path length of \SI{5}{\milli\meter}. In black: transmission spectra for the UNIPLAST slide with light path lengths of \SI{7}{\milli\meter} (solid line) and \SI{35}{\milli\meter} , respectively (dashed line: at sample position 1, dotted line: at sample position 2, as described in the text).
    } 
    \label{fig:Transmission_spectra}
\end{figure}

\begin{figure}[!htbp]
    \centering
    \includegraphics[width=0.8\textwidth]{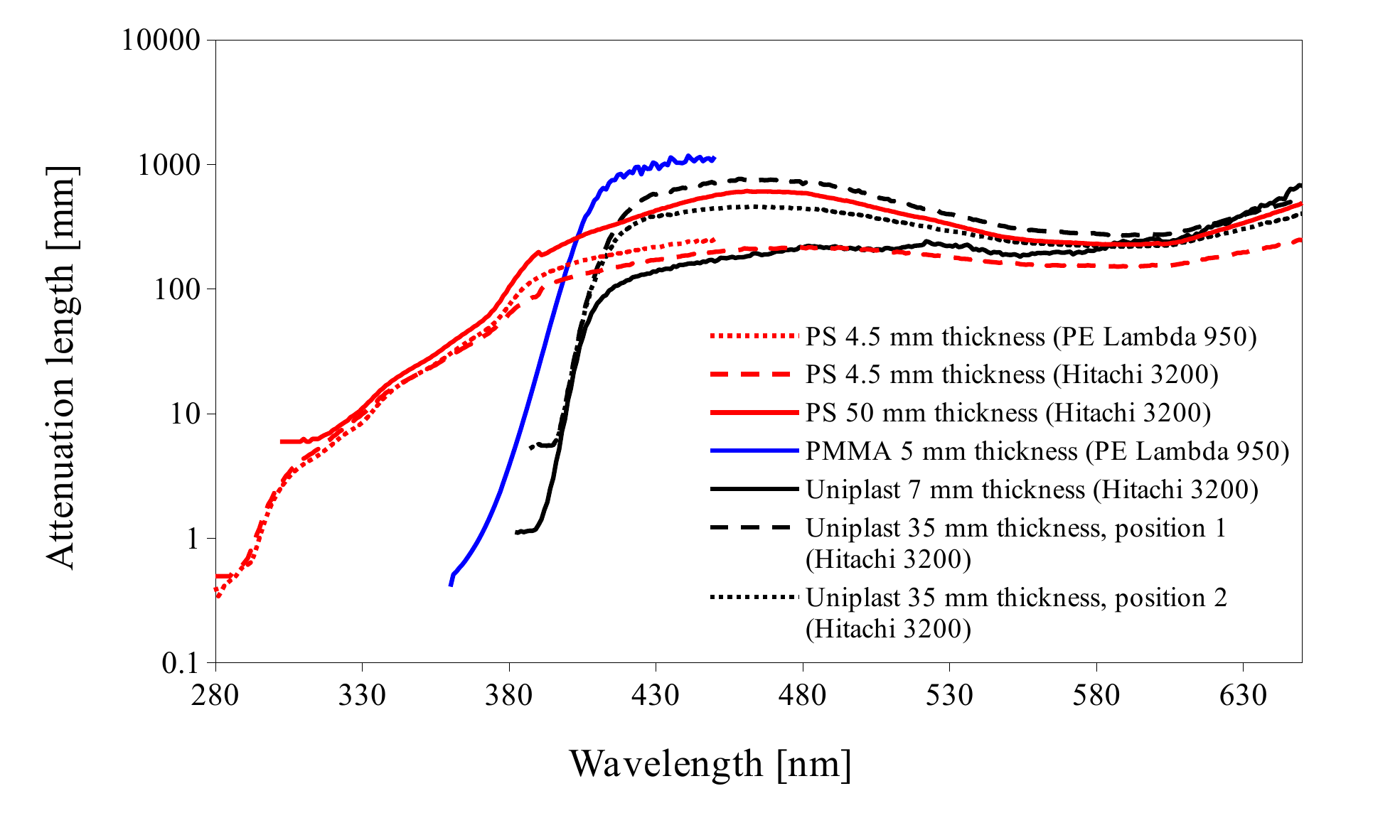}
    \caption{Attenuation lengths extracted from the transmission spectra presented in Fig.\,\ref{fig:Transmission_spectra} for samples made of polystyrene (PS), PMMA, and UNIPLAST. The meaning of the line styles are identical to those in Fig.\,\ref{fig:Transmission_spectra}.
    The attenuation lengths are only plotted in those wavelengths ranges where the spectrophotometers have sufficient sensitivity
    to measure transmission values above the noise level.
    } 
    \label{fig:AttenuationLength_spectra}
\end{figure}

Transmission spectra as a function of wavelength were measured for various samples with two different transmission spectrophotometers, using A) a PerkinElmer Lambda 950 spectrophotometer (PE Lambda 950) measuring in the range between \SI{280}{\nano\meter} and \SI{450}{\nano\meter}, B) a Hitachi 3200 spectrophotometer (Hitachi 3200) \cite{HITACHI-U-3200} measuring in the range between \SI{280}{\nano\meter} and \SI{900}{\nano\meter}. The attenuation length as a function of wavelength was calculated by correcting the measured transmission $T_{meas}$ for the reflectivity $R$ of light at the front and back surface of the sample, taking into account multiple reflections\cite{scharf:2018rad}:

\begin{equation}
    \tag{1}
    \label{Eq:Att_Length}
    \lambda_{abs} = \frac{d}{\text{ln}\Biggl(\cfrac{T^2 + \sqrt{T^4 + 4 R^2 T_{meas}^2}}{2 T_{meas}}\Biggr)} 
\end{equation}

 where $T$ and $R$ are the predicted transmittance and reflectivity for the case of no absorption in the material calculated as 
\begin{equation}
    \tag{2}
    R = 1 - T = \left|\frac{1 - n}{1 + n}\right|^2
\end{equation}
with $n$ the wavelength-dependent refraction index of the material taken from \cite{refr_index}.
An overview of the studied material samples is provided in
Table \,\ref{Tab:TransmissionSamples}.
The corresponding transmission measurements are summarized in Fig.\,\ref{fig:Transmission_spectra}
and the corresponding attenuation lengths
in Fig.\,\ref{fig:AttenuationLength_spectra}.

\begin{table}[!h]
    \centering
    \caption{List of plastic material samples (pure PS, PMMA, and UNIPLAST extruded scintillator) used for the transmission measurements. The geometrical dimensions of the planar slide geometry are quoted as well as the light path length through the material, and the spectrophotometer used for the transmission measurement. For the UNIPLAST scintillator, two different beam positions on the sample
    were measured to quantify the systematic uncertainty
    on the transmission measurement and corresponding attenuation length as a result of the inhomogeneous material transparency.}
    \vspace{1.2mm}
    \begin{tabular}{cccc}
        \hline\hline
Slide    & Dimension  & Light path length & Spectro-\\
Material &${\rm mm}^3$& mm        &  photometer\\
        \hline
PS       & $50 \times 50 \times 4.5$ & 4.5 & PE Lambda 950 \\
PS       & $50 \times 50 \times 4.5$ & 4.5 & Hitachi 3200 \\
PS       & $50 \times 50 \times 4.5$ & 50  & Hitachi 3200 \\
        \hline
PMMA     & $50 \times 50 \times 5.0$ & 5.0 & PE Lambda 950 \\
        \hline
UNIPLAST & $35 \times 35 \times 7$ &  7  & Hitachi 3200 \\ 
UNIPLAST & $35 \times 35 \times 7$ & 35  & Hitachi 3200 \\ 
         &                         & position 1  &  \\
UNIPLAST & $35 \times 35 \times 7$ & 35  & Hitachi 3200 \\ 
         &                         & position 2  &  \\
        \hline
        \hline
    \end{tabular}
    \label{Tab:TransmissionSamples}
\end{table} 

At small wavelengths, where in general the attenuation length becomes significantly smaller than the actual sample thickness, the spectrophotometer transmission measurements lose sensitivity due to the very small light intensities. As a result, we only plot the extracted attenuation lengths 
above 
these wavelength values.
 
The transport of the UV photons, either generated by scintillation or by Cherenkov radiation, to the WLS coating requires a sufficiently large transparency (attenuation length) of the plastic material in the UV range \SIrange{280}{400}{\nano\meter} to avoid
too large absorption losses before reaching the WLS layer on the slide surface.
Secondary photon emission from the WLS
layer in the visible range between \SI{400}{\nano\meter}
and \SI{500}{\nano\meter}, where many types of photomultipliers
or silicon photomultipliers reach also
the highest sensitivities, requires even
larger attenuation lengths in this
wavelength range to build detectors
of larger volume.

Fig.\,\ref{fig:Transmission_spectra} shows the transmission spectrum through the uncoated, 5 mm thick PMMA slide (blue line) and
Fig.\,\ref{fig:AttenuationLength_spectra} the corresponding attenuation length. The PMMA material under study becomes only transparent above about \SI{380}{\nano\meter}.
As a result, only Cherenkov photons within the range \SIrange{380}{400}{\nano\meter} can be absorbed by the WLS layer on the surface of the PMMA slide and therefore a rather small light yield is to be expected from generation of Cherenkov light by charged particles traversing this material.

Fig.\,\ref{fig:Transmission_spectra} shows as well the transmission spectra for the pure PS slide (red dotted and dashed lines) for a light path length through the material of about \SI{4.5}{\milli\meter} measured with both spectrophotometers. The results obtained with both spectrophotometers are in very good agreement for wavelengths below \SI{380}{\nano\meter}. 

In contrast to the PMMA material, the PS material shows transparency already at about \SI{290}{\nano\meter} (see Fig.\,\ref{fig:Transmission_spectra}) with an attenuation length close to \SI{10}{\milli\meter} at \SI{325}{\nano\meter} (see Fig.\,\ref{fig:AttenuationLength_spectra}), where the emission maximum of scintillation light in PS is located. As a result, much more of the UV photons that are generated in the PS slide can be absorbed in the WLS compared to the PMMA slide.

The optical transmission of the $50 \times 50 \times 4.5 {\rm mm}^3$ PS sample was measured as well with the Hitachi 3200 spectrophotometer for a \SI{50}{\milli\meter} light path length through the material, which allows to determine the attenuation length of the PS material also with good precision at higher wavelengths where the attenuation length is much larger than \SI{4.5}{\milli\meter} thickness of the slide. 
The corresponding transmission curves measured up to wavelenghts of 650 nm are summarized in Fig.\,\ref{fig:Transmission_spectra} and the corresponding attenuation lengths in Fig.\,\ref{fig:AttenuationLength_spectra}. Wavelengths well above 500 nm are not relevant for our study because of the emission spectra of the materials under study and the fact that the R1924A PMT used in our studies with a $\beta$ source (see section \ref{Sec:ExperimentalSetup}) has negligible quantum efficiency, respectively, sensitivity above 650 nm.

Since the transmission results for the
\SI{4.5}{\milli\meter} PS slide 
measured with both spectrophotometers are in
(very) good agreement, the attenuation length results are very similar for both spectrophotometer results as well.
The attenuation length result obtained for the \SI{50}{\milli\meter} light path length agrees with the two results obtained for \SI{4.5}{\milli\meter} light path length in the wavelength range between \SI{300}{\nano\meter} and \SI{360}{\nano\meter}. Hence, we consider the results from \SI{50}{\milli\meter} thickness valid as well for larger wavelengths above \SI{360}{\nano\meter}, where the results for the \SI{4.5}{\milli\meter} thickness are less trustworthy.
The attenuation length results for the \SI{50}{\milli\meter} thickness can be considered trustworthy as long as the extracted attenuation length does not become too large with respect to the sample thickness, as for example in the region above  \SI{630}{\nano\meter}.
In the WLS emission range \SIrange{400}{500}{\nano\meter}, the
attenuation length extracted from the \SI{50}{\milli\meter} light path length
measurement grows from \SI{20}{\centi\meter} to \SI{50}{\centi\meter}.

The obtained results of the PS attenuation length has consequences for building a WLS-coated PS plastic scintillator detector:
\begin{enumerate}
\item As long as the plastic scintillator detector thickness is not much larger than the attenuation length
of about \SI{7}{\milli\meter} at \SI{320}{\nano\meter}, which corresponds to the emission maximum of primary scintillation light in pure PS, a large fraction of these primary UV photons can be absorbed by the wavelength shifter in the thin layer on the slide surface. Hence, a slide thickness of about \SI{5}{\milli\meter} is well chosen, although the thickness is subject to optimization because the number of primary scintillation photons produced by a traversing charged particle increases with the scintillator thickness.
\item  Due to the attenuation length of the order \SI{10}{\milli\meter} in the UV range, the determination of the primary scintillation light yield requires setups using rather small PS sample sizes, as described in \ref{Sec:ScintillationLightYieldPurePS}.
\item  To build longer strips made of WLS-coated PS plastic scintillator material, with photon detection taking place at the small cross-section ends of the long strip, the attenuation length in the visible range (\SI{400}{\nano\meter} - \SI{500}{\nano\meter}) and in particular close to the WLS emission maximum of \SI{425}{\nano\meter} is relevant. The attenuation length results for the pure PS material suggest that scintillator strips with sufficient light-yield output on both strip sides of up to \SI{50}{\centi\meter} might be possible, which will be studied in more detail in future work.
\end{enumerate}

The transmission through a UNIPLAST slide was measured as well with the Hitachi 3200 spectrophotometer (see Fig.\,\ref{fig:Transmission_spectra}) with the corresponding attenuation-length results shown in Fig.\,\ref{fig:AttenuationLength_spectra},
using a $35 \times 35 \times 7 {\rm mm}^3$  sample, with two different orientations, resulting in light path lengths 
through the material of either 7 mm or 35 mm. 
Since the UNIPLAST material is a PS-based scintillator, it is assumed
that the extruded UNIPLAST scintillator has the same refraction index as pure polystyrene, when translating the transmission spectrum into the attenuation length as function of wavelength 
(see Fig. \,\ref{fig:AttenuationLength_spectra}). This assumption is supported by the fact that the transmission spectra
for UNIPLAST and pure PS agree reasonably well at larger wavelengths
(above about 480 nm) for the sample thickness of 7 mm (UNIPLAST) respectively 4.5 mm (PS) where the transmission is dominated by the reflection of light at the front and back side of the slides.

We note that the UNIPLAST scintillator material has a visible non-uniformity in transparency inside the samples, which generates significant differences in the measured transmission spectra and hence the corresponding attenuation lengths when choosing different intersection points for the same light path length through the sample. Therefore, the transmission-spectrum results for two
different positions (called 1 and 2) of the light beam on the UNIPLAST slide for the \SI{35}{\milli\meter} light path length are presented.

\subsubsection{Scintillation light yield of pure polystyrene and of the UNIPLAST reference samples}
\label{Sec:ScintillationLightYieldPurePS}

\begin{figure}[!htbp]
    \centering
    \includegraphics[width=0.9\linewidth]{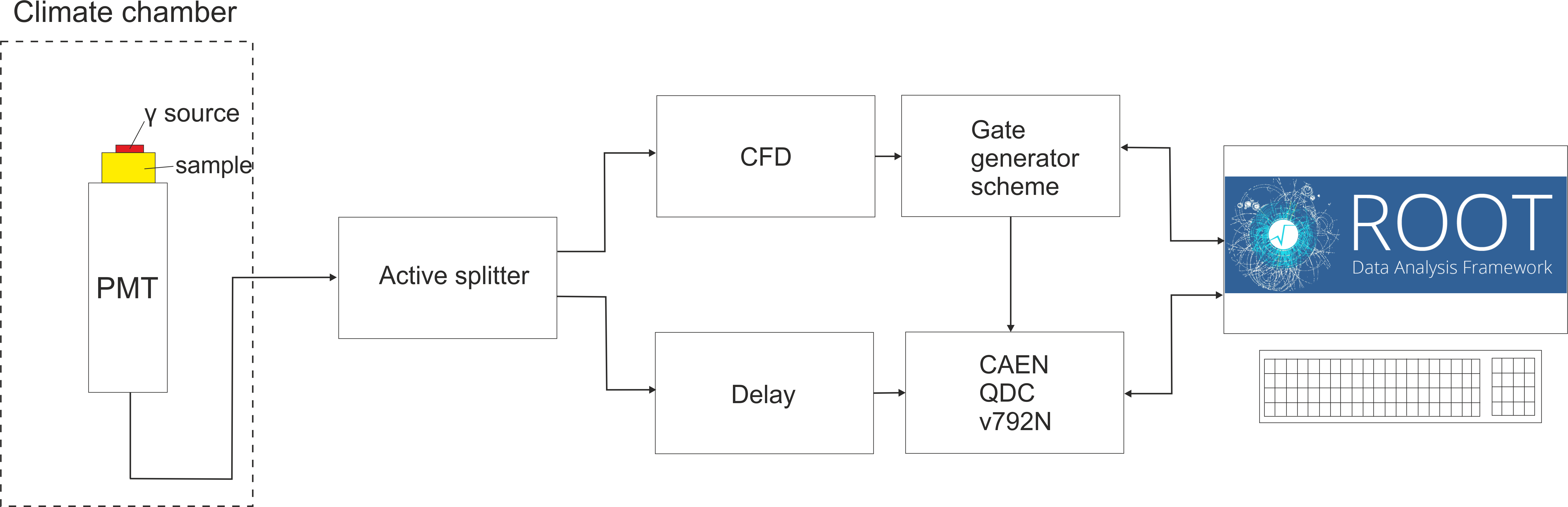} 
    \caption{Scheme of the setup for light yield measurements. A signal from the  R2059-01 PMT is splitted in two. One goes through the passive delay chain to an input of a CAEN QDC v792N unit, the second goes through a constant fraction discriminator and forms the integrating gate for the QDC. The data acquisition is performed with a PC based on the ROOT software.
    } 
    \label{fig:LY_setup}
\end{figure}

\begin{figure}[!htbp]
    \centering
    \includegraphics[width=0.9\linewidth]{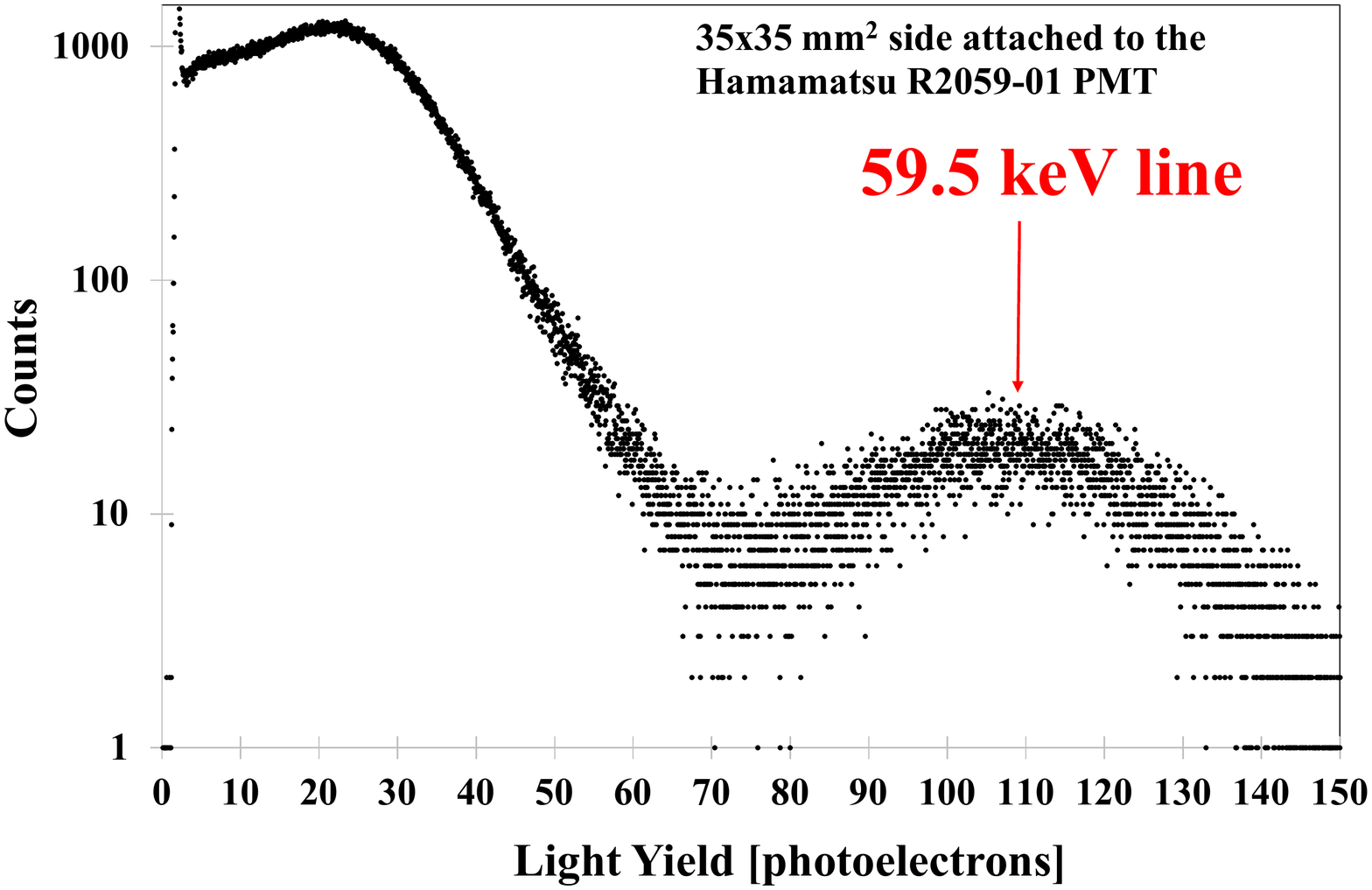} 
    \caption{Photoelectron yields obtained with a UNIPLAST sample of dimensions $35 \times 35 \times 7 {\rm mm}^3$ coupled to a calibrated reference PMT Hamamatsu R2059-01
    using a $^{241}$Am ($E_{\gamma}$ = \SI{59.5}{\kilo\electronvolt}) source.  The acquisition time of every spectrum was 600 seconds, the integration time gate was 1000 ns.
    } \label{fig:UNIPLAST_Am241}
\end{figure}

\begin{figure}[!htbp]
    \centering
    \includegraphics[width=0.9\linewidth]{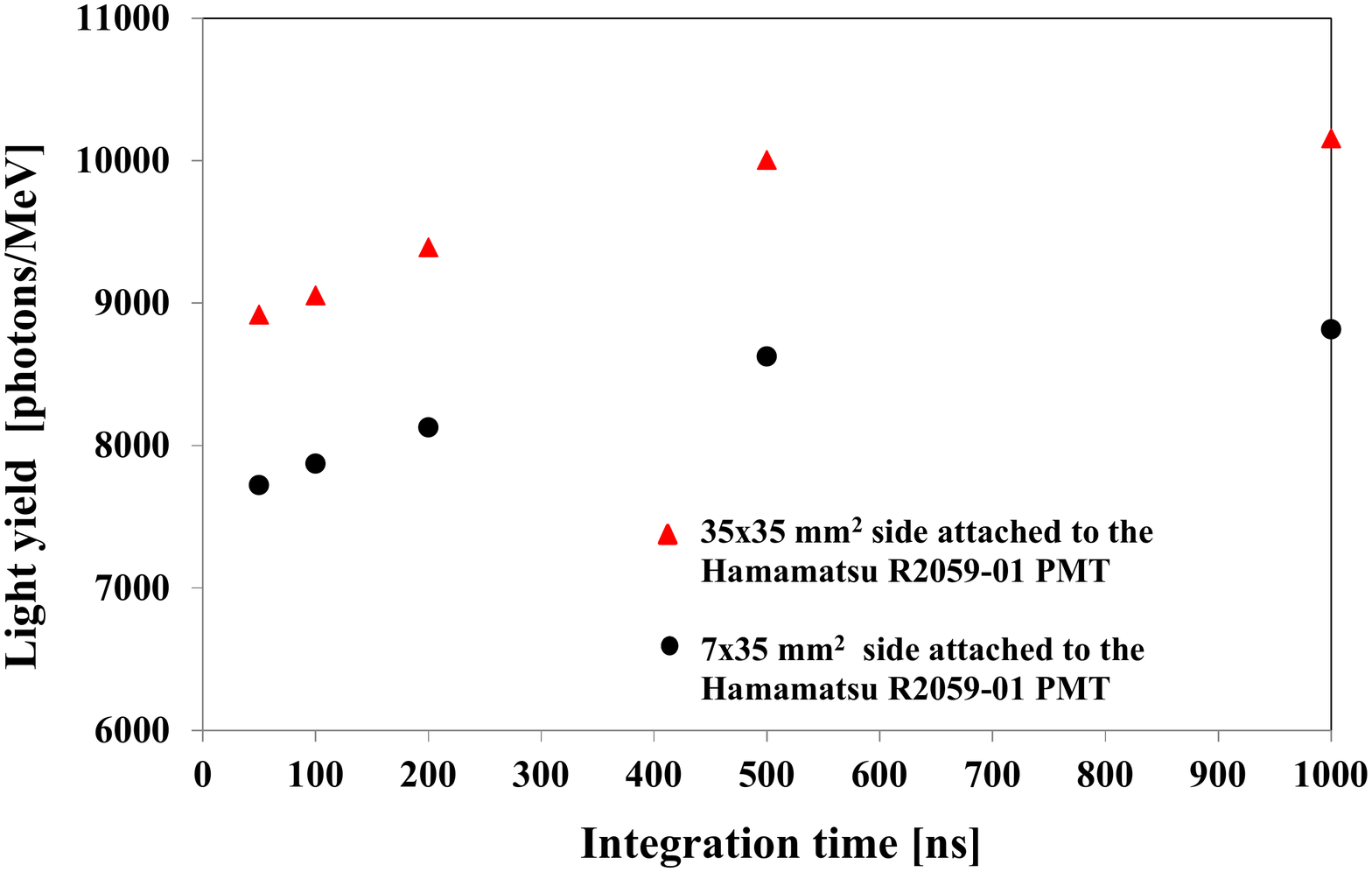} 
    \caption{Scintillation light yield in photons per MeV energy deposition as a function of integration time of a $35 \times 35 \times 7 {\rm mm}^3$ UNIPLAST sample, with two different sides coupled to
    the calibrated R2059-01 PMT as described in the text.
    } \label{fig:ScintLightYield_IntegrationTime_UNIPLAST}
\end{figure}

Using a Hamamatsu R2059-01 \cite{Hamamatsu-R2059-01} reference PMT with a calibrated bialkali photocathode, the light yield of a UNIPLAST sample with dimensions of $35 \times 35 \times 7 {\rm mm}^3$ was measured at room temperature
as a function of integration time. The scheme of the setup for the light yield measurement is shown in Figure \ref{fig:LY_setup}. All sample surfaces were optically polished and the sample was wrapped inside multiple layers of Teflon film serving as a diffuse light reflector. Baysilone oil M 300.000 (Momentive Performance Materials Company) was used as an optical coupling agent to couple the UNIPLAST scintillator material to the R2059-01 PMT. The position of the PMT's single electron peak was used for the light yield value calibration \cite{Lecoq:1995ly, Zhu:1996ly}. The response to an $^{241}$Am ($E_{\gamma}$ = \SI{59.5}{\kilo\electronvolt}) $\gamma$-source was measured for two different orientations of the sample on the R2059-01 PMT entrance window.  The source was placed on the sample side opposite to the side attached to the PMT. A noise contribution is measured without  a sample on the PMT for every integrating gate as a pedestal value and subtracted from the values measured with the samples and the source installed on the PMT. The $^{241}$Am spectrum  is shown in Figure \ref{fig:UNIPLAST_Am241} as example. Since we used a calibrated PMT, one can easily convert the light yield measured in photoelectrons/MeV into photons/MeV taking into account a quantum efficiency at the luminescence maximum of the material as QE(420 nm) = 17.7\%. The results of the light yield measurement are summarized in Figure \ref{fig:ScintLightYield_IntegrationTime_UNIPLAST}. One observes that the “horizontal” position ($35 \times 35 {\rm mm}^2$ side attached to R2059-01 PMT) results in about 12 \% higher value of the light yield in comparison with the “vertical” position ($7 \times 35 {\rm mm}^2$ side attached to PMT). This can be explained by the fact that for the “vertical” position the average path length of scintillating photons is larger. The absolute value of the light yield reaches about \SI{10000}{ph\per\mega\electronvolt} for a 1 microsecond of integration time.

\begin{figure}[!htbp]
    \centering
    \includegraphics[width=0.9\linewidth]{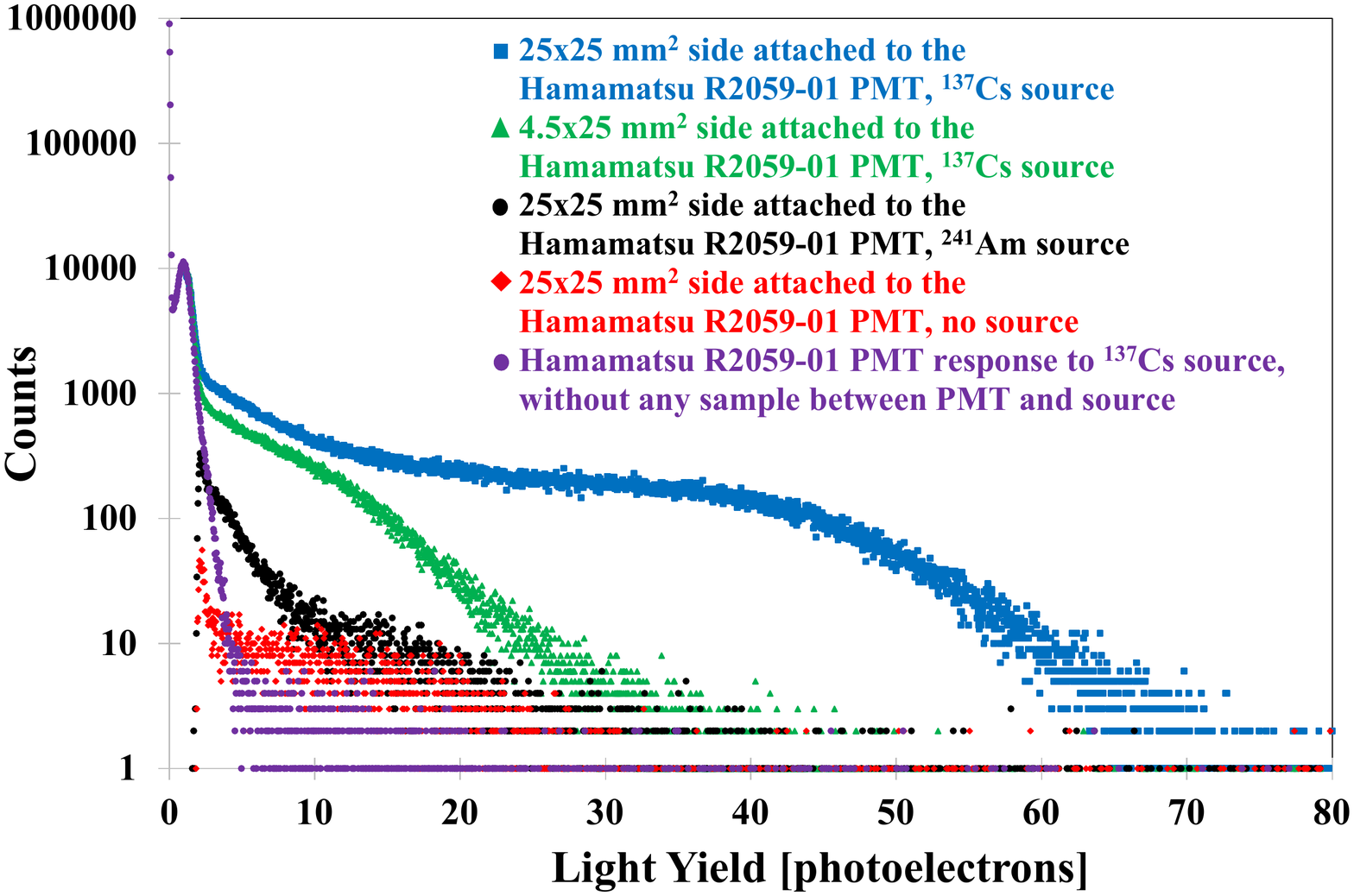} 
    \caption{Photoelectron yields obtained with a pure polystyrene sample of dimensions $25 \times 25 \times 4.5 {\rm mm}^3$ coupled to a calibrated reference PMT Hamamatsu R2059-01
    using a $^{137}$Cs ($E_{\gamma}$ = \SI{662}{\kilo\electronvolt}) $\gamma$-source as described in the text.
    Horizontal position on PMT (blue): $25 \times 25{\rm mm}^2$ side attached to the PMT. Vertical position (green): $4.5 \times 25{\rm mm}^2$ side attached to the PMT.
    Photoelectron yield  without any source (red) and with a $^{241}$Am ($E_{\gamma}$ = \SI{59.5}{\kilo\electronvolt}) $\gamma$-source (black), single electron peak and a response of the PMT without a sample on the $^{137}$Cs source (violet). For all spectra excepting the last, a threshold cutting the single electron peak was applied. The acquisition time of every spectrum was 600 seconds, the integration time gate was 1000 ns.
    } 
    \label{fig:LightYield_log}
\end{figure}

The light yield of samples of pure polystyrene were characterized as well
with the R2059-01 reference PMT. The samples have dimensions of $50 \times 50 \times 4.5 {\rm mm}^3$ and $25 \times 25 \times 4.5 {\rm mm}^3$. 
To minimize the effect of the photon absorption due to long path lengths, we tested the pure PS sample with dimensions of $25 \times 25 \times 4.5 {\rm mm}^3$
with a $^{241}$Am
$\gamma$-source as well
as with a $^{137}$Cs $\gamma$-source.
The measurements were done at the same conditions 
as for the UNIPLAST sample and for two orientations on the R2059-01 PMT: “horizontal” ($25 \times 25 \,{\rm mm}^2$ side attached to the PMT) and “vertical” ($4 \times 25{\rm mm}^2$ side attached to the PMT). The results are presented in Figure 
\ref{fig:LightYield_log}.

No well-identifiable response was observed with a $^{241}$Am $\gamma$-source for both sample orientations,
 as shown in Fig. \ref{fig:LightYield_log}.
No photo peak of \SI{662}{\kilo\electronvolt} $\gamma$ quanta was observed, likely due to the low density of polystyrene and the relatively small thickness of the sample, but one can clearly recognize the shift of the Compton edge to lower photoelectron (PE) yield when changing from the “horizontal” to the “vertical” orientation of the sample.
Qualitatively, this behaviour is to be expected:
in vertical position, the Compton scattering positions are distributed over a length of \SI{25}{\milli\meter}  compared to only \SI{4.5}{\milli\meter}  in the case of horizontal position.
Since the attenuation length in the UV range for pure PS is rather small (around \SI{10}{\milli\meter} at \SI{320}{\nano\meter} wavelength, see Fig. \ref{fig:AttenuationLength_spectra}), one expects to collect less photons in vertical position.
The position of the Compton edge for the “horizontal" orientation is located at about 60 photoelectrons and corresponds to 2/3 of the $\gamma$ energy (\SI{441}{\kilo\electronvolt} for \SI{662}{\kilo\electronvolt} $\gamma$ quanta). Using the calibrated PMT Hamamatsu R2059-01, we convert the light yield measured in photoelectrons/MeV into photons/MeV, taking into account a quantum efficiency at the luminescence maximum of polystyrene as QE(325 nm) = 24.22\% and the known energy deposition of the \SI{441}{\kilo\electronvolt} electron
(corresponding to the Compton edge in the photoelectron spectrum). From the 60 photoelectrons observed at the Compton edge
when attaching the sample with the $25 \times 25{\rm mm}^2$ side to the PMT, one obtains
$60/0.2422/0.441 = 560$ ± (5\%) photons/MeV.
This light yield is likely still underestimated due to the fact that the attenuation length inside pure PS is not much larger than the sample dimensions (see Fig. \ref{fig:AttenuationLength_spectra}). 

\subsection{Measurements with the $\beta$-source setup}
\label{Sec:ExperimentalSetup}

This sub-section decribes the setup to measure the number of photoelectrons detected with
the coated PS and PMMA slides and the UNIPLAST reference slide coupled always to a Hamamatsu-R1924A PMT\cite{Hamamatsu-R1924A} and using the same $^{90}$Sr $\beta^{-}$ source. We present first the calibration of the R1924A PMT used in our dedicated
$\beta^{-}$ setup.

\subsubsection{Calibration of the R1924A PMT with a dedicated laser setup}
\label{Sec:CalibrationLaserSetup}

\begin{figure}[!htbp]
    \centering
    \includegraphics[width=0.9\linewidth]{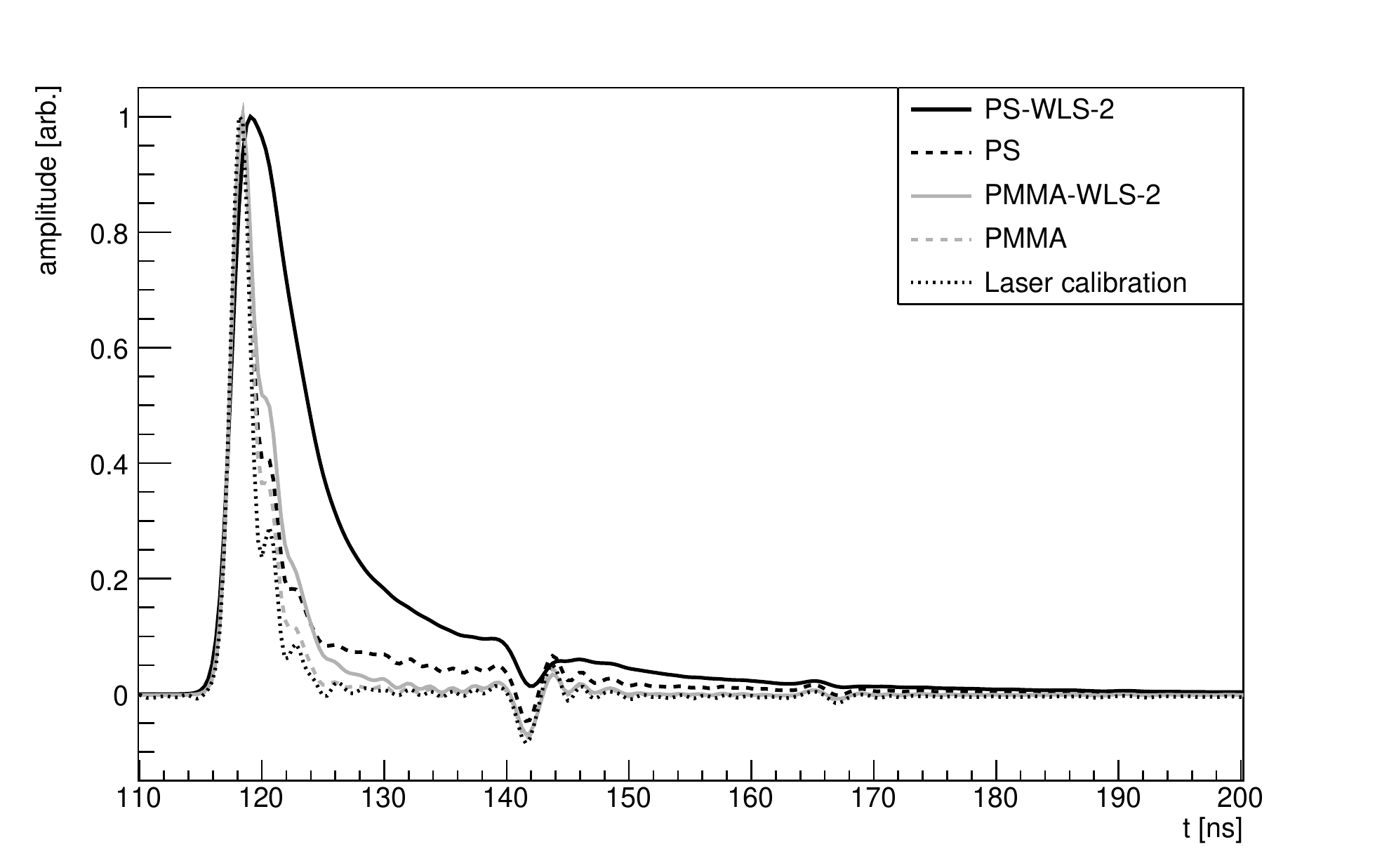}

    \caption{
    The averaged waveform (dotted line), normalized to the maximum amplitude, for the calibration of the R1924A PMT with the laser setup is shown, where about 3 photoelectrons on average are detected by the PMT within $\text{FWHM}=\SI{54}{\pico\second}$. The other curves show a comparison of the averaged waveforms normalized to the maximum amplitude for the uncoated PS (black dashed line) and PMMA (grey dashed line) slides and the double coated PS-WLS-2 (black solid line) and the PMMA-WLS-2 (grey solid line) slides, the latter of which are very similar to the corresponding single coated PS-WLS and PMMA-WLS slides (not shown). 
    } 
    \label{fig:SumOfWaveFormsMeasurements}
\end{figure}

\begin{figure}[!htbp]
    \centering
    \subfloat[1.8 PE per laser pulse]{
    \includegraphics[width=0.45\textwidth]{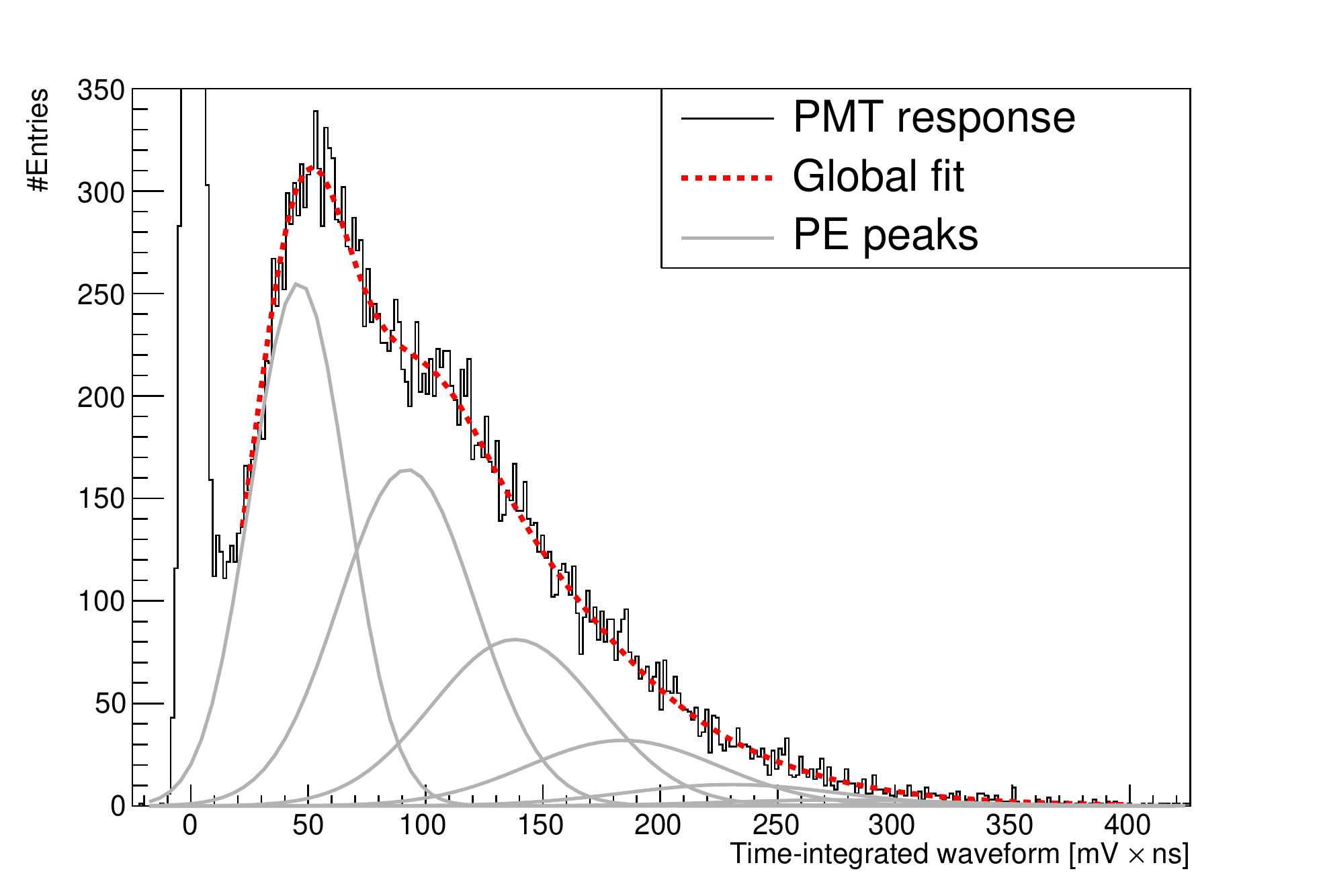}
        }
    ~ 
    \subfloat[2.5 PE per laser pulse]{
        \includegraphics[width=0.45\textwidth]{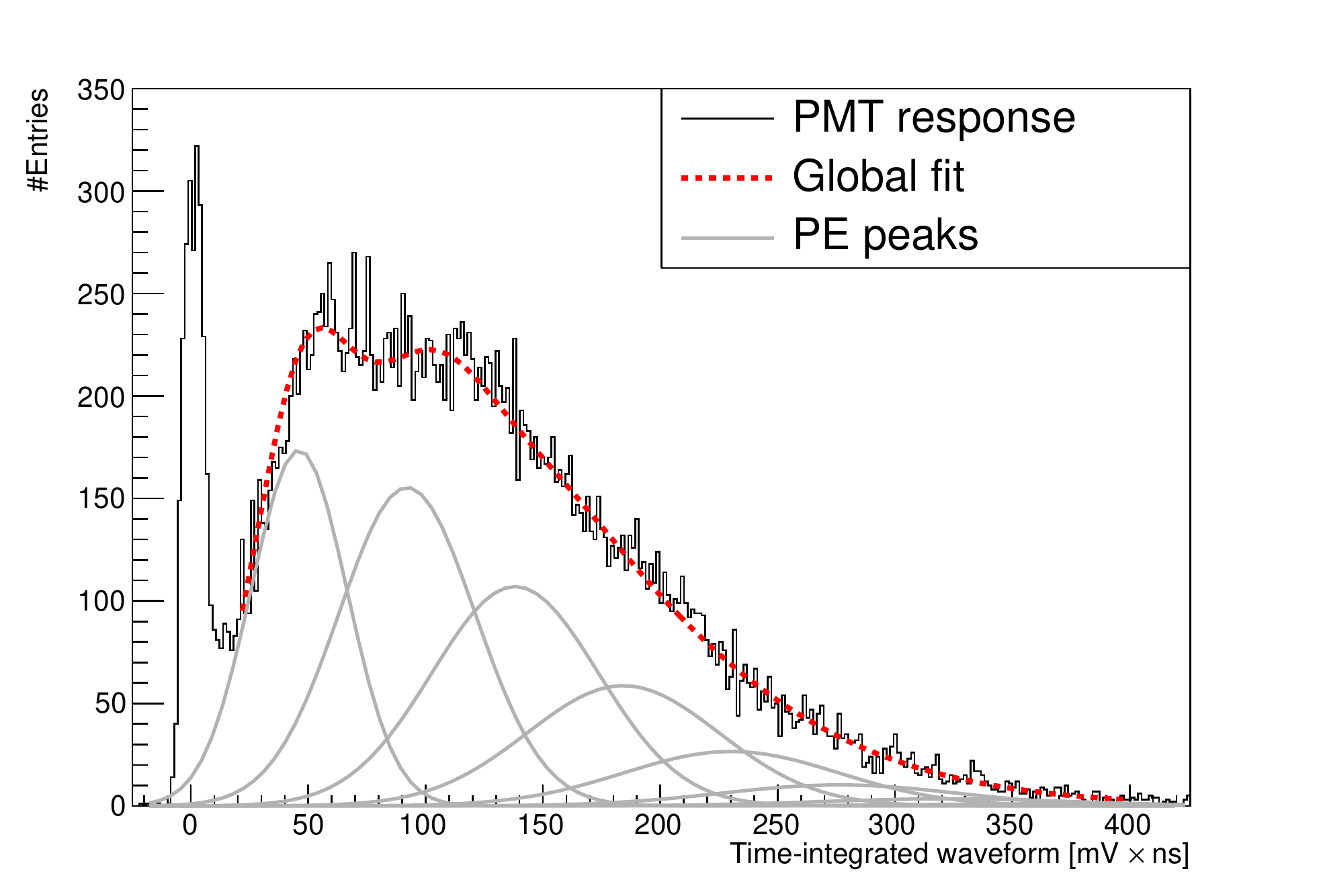}
        }
    
    \subfloat[2.9 PE per laser pulse]{
        \includegraphics[width=0.45\textwidth]{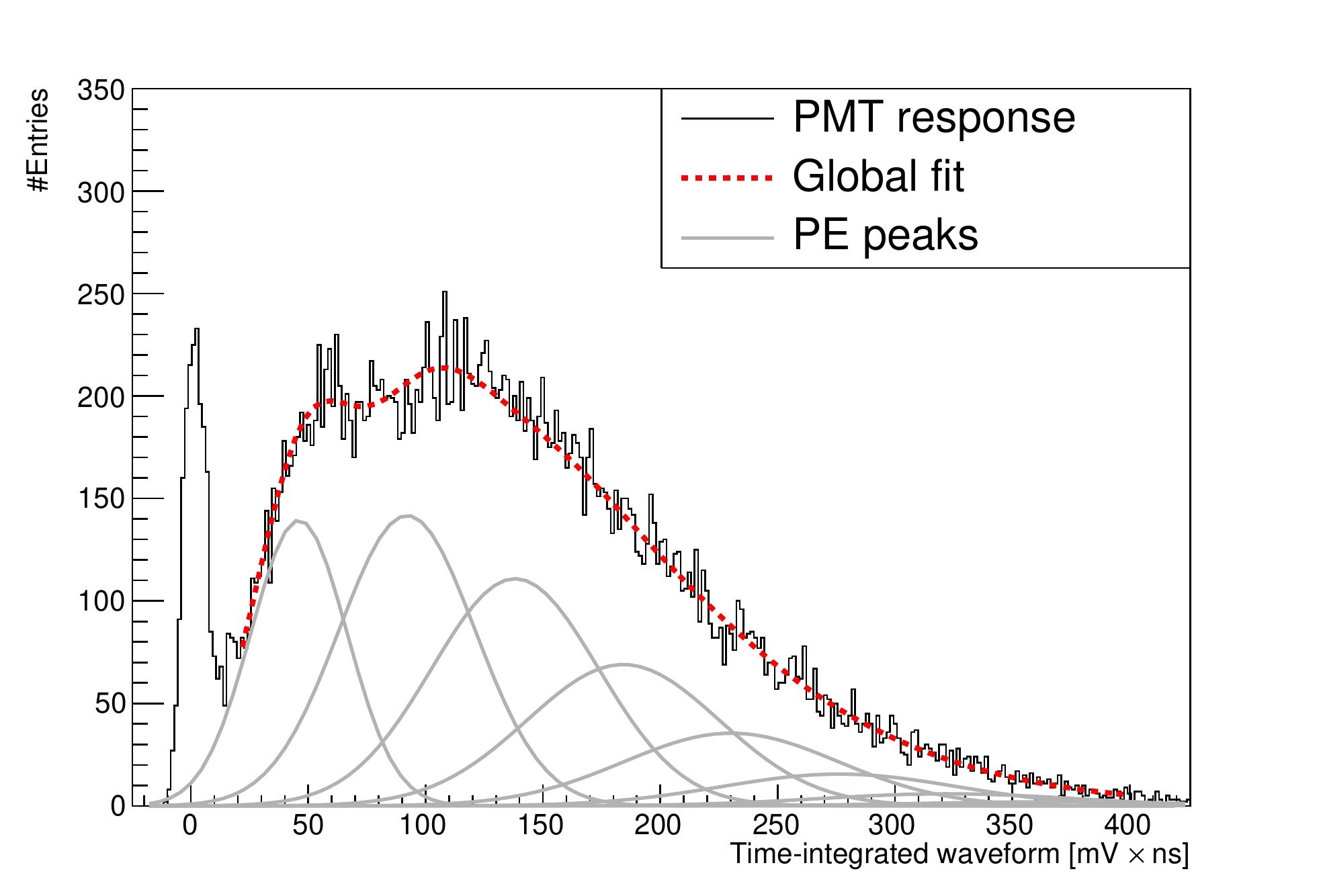}
        }
    \caption{Spectra of the time-integrated waveforms for the gain calibration measurements of the R1924A PMT at a bias voltage of \SI{1000}{\volt} using short laser pulses that are reduced in intensity by attenuators resulting in a low
    average number of detected PE in the PMT per pulse (average number of PE given in the captions, compare Fig.\,\ref{fig:SumOfWaveFormsMeasurements} for corresponding waveform forms). The time-integrated spectra (black lines) are fitted simultaneously with the fit function described in the text. The gain value extracted from the global fit (dashed red lines) is used to translate time-integrated spectra into PE spectra. The individual PE peaks (gray lines) from the global fit are also shown.}
    \label{fig:TimeIntegratedCalibration}
\end{figure}

To quantify the PE yield detected by the R1924A PMT, measurements with a pulsed laser (PILas\,\cite{PILas} delivering pulses with a full width at half maximum of \SI{54}{\pico\second} at a wavelength of \SI{402.6}{\nano\meter}) were performed. The PMT was placed inside a dedicated dark box and the laser light pulses were directed through a diffusor onto the PMT cathode, not using the plastic slides. 
The PMT signals were read out using a 16-channel WaveCatcher digitizer \cite{WaveCatcher}. The data were collected in runs, consisting of a set of events defined by the measurement trigger.
Data taking for all these measurements was triggered with the trigger signal from the laser driver. 
Each run was then analysed using a C++ code based on the ROOT software framework \cite{Antcheva:2011zz}.
Fig.\,\ref{fig:SumOfWaveFormsMeasurements} shows the averaged recorded waveforms normalized to the maximum amplitude for a laser calibration measurement.
The waveforms of each individual event for these measurements were time-integrated within an integration window of \SI{16}{\nano\second} width, starting \SI{4}{\nano\second} before the maximum of the signal. This time window was chosen to be significantly shorter compared to the measurements with the plastic slides to minimize the effect from noise. Resulting time-integrated spectra are shown in Fig.\,\ref{fig:TimeIntegratedCalibration}.

The gain of the R1924A PMT is determined from a fit of the ideal PMT function (Eq.\,5 of\,\cite{Bellamy:1994bv}) to the time-integrated spectra, which allows one to translate the time-integral into number of photoelectrons. Since the gain of the R1924A PMT is comparable to two times the standard deviation of the PE peaks, it is difficult to distinguish the individual peaks and to determine the gain. Under these circumstances, the individual peaks are best distinguishable if the mean number of 
PE detected by the R1924A PMT per laser pulse is between 1.5 and 3.0 PE. For the calibration, we have tuned the laser intensity such that we achieve a mean number of PE in this range and have taken three measurements at different mean number of PE. The three measurements are fitted simultaneously with the gain and the standard deviation of the PE peaks as global fit parameters.

The calibration measurements are shown in Fig.\,\ref{fig:TimeIntegratedCalibration} for a PMT bias voltage of 1000\,V. The global fit of the laser calibration measurements gives 1 PE = (46.05 $\pm$ 0.47) mV·ns for a bias voltage of 1000\,V and 1 PE = (19.97 $\pm$ 0.21) mV·ns for a bias voltage of 900\,V.

\subsubsection{Measurements with the dedicated $\beta$-source setup}
\label{Sec:betaSetup}

\begin{figure}[b]  \centering   \includegraphics[width=0.6\textwidth]{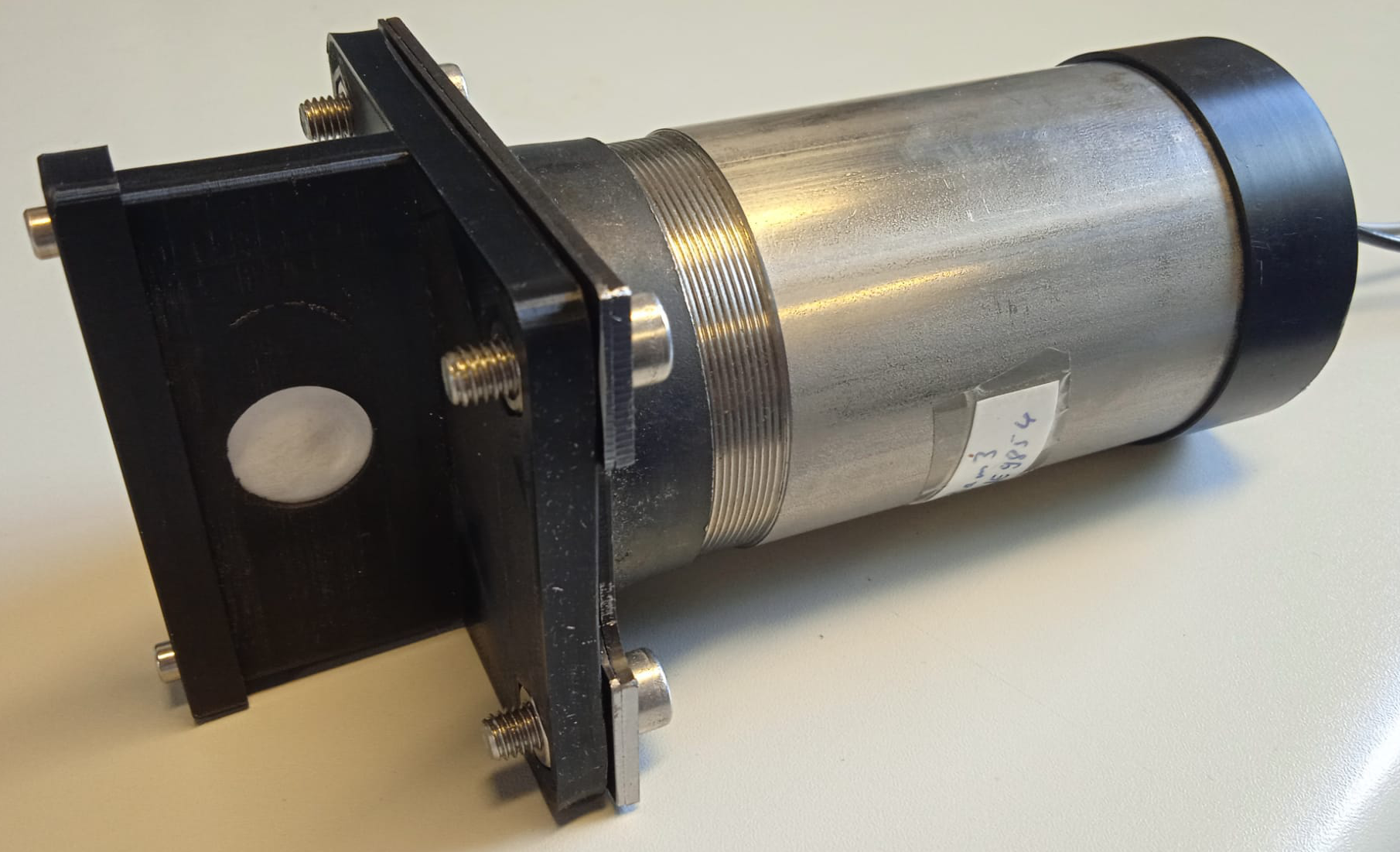}
    \caption{The plastic slide, wrapped in Dupont\texttrademark \hspace{0.5mm} Tyvek\textsuperscript{\textregistered} 1073D paper, enclosed inside a black plastic cover, coupled to the R1924A PMT, which is surrounded by a metal housing.
    } 
    \label{fig:PlatePlusPMT}
\end{figure}

To ensure a well-defined position of the plastic slide under study, it was placed inside a black plastic cover of \SI{1.2}{\milli\meter} wall thickness. 
To enable high-energetic $\beta^{-}$ particles traversing the plastic material slides, the cover had circular holes of \SI{16.8}{\milli\meter} diameter on the top and bottom side that were covered with thin black tape to guarantee minimal energy loss together with light tightness. 
The slide was then coupled to one of the four \SI{5}{\milli\meter} sides via optical gel (Bayer silicone Baysilone) to the Hamamatsu R1924A PMT (cathode diameter: \SI{22}{\milli\meter}, quantum efficiency QE = \SI{26}{\percent} at \SI{420}{\nano\meter}) \cite{Hamamatsu-R1924A}, as shown in Fig.\,\ref{fig:PlatePlusPMT}, operated at a voltage of \SI{1000}{\volt}.

To increase the light collection efficiency, 
the plastic slides are wrapped with Dupont\texttrademark \hspace{0.5mm} Tyvek\textsuperscript{\textregistered} 1073D paper, a diffuse reflector, only leaving out \SI{10}{\milli\meter} of the side in contact with the R1924A PMT to prevent the optical gel from leaking between the slide and the wrapping.

As for the calibration measurements, the R1924APMT signals were read out using the same 16-channel WaveCatcher digitizer. The trigger signal was built from two small circular scintillator slides placed right on top of each other, each coupled using optical gel (Bayer silicone Baysilone) to a Hamamatsu R5900 PMT operated at a voltage of \SI{780}{\volt} and installed inside a box ("trigger box"). The upper scintillator has a diameter of \SI{20}{\milli\meter} diameter, the lower one has a diameter of \SI{18}{\milli\meter}, and both of them are \SI{1}{\milli\meter} thick. The trigger box was located \SI{3}{\milli\meter} below the slide under study.

\begin{figure}[]  
    \centering  \includegraphics[width=0.45\textwidth]{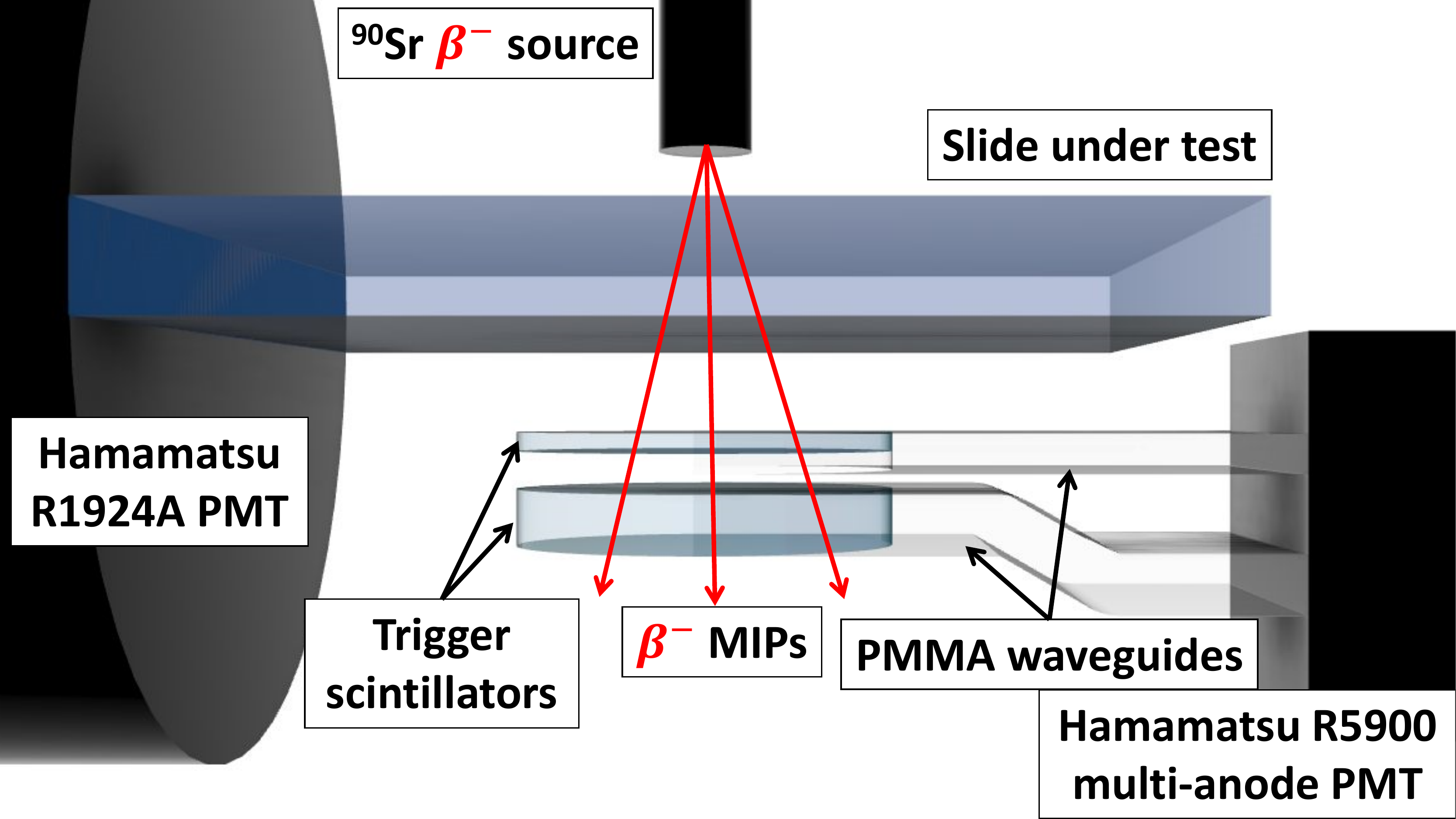}
    \caption{The experimental setup consists of a $^{90}$Sr source located above the plastic slide under study coupled to the R1924A PMT and the trigger box with the two trigger scintillators, each coupled to a Hamamatsu R5900 PMT, located inside the trigger box below the plastic slide.
    } 
    \label{fig:SetupSketch}
\end{figure}

Data were taken with a $^{90}$Sr source located \SI{4}{\milli\meter} above the plastic slide, as shown in Fig.\,\ref{fig:SetupSketch}.
Given the thickness of the plastic slides under study and
the other materials to be traversed by the electrons from the $\beta^{-}$ source
to reach the lower trigger scintillator, the $\beta^{-}$ kinetic energy required is
at least \SI{2}{\mega\electronvolt}. This condition is fulfilled by a fraction of electrons
coming from $\beta^{-}$ decays of the $^{90}$Sr daughter isotope $^{90}$Y. 
The trigger rate, within the trigger window of 15 ns, without the $^{90}$Sr source was \SI{0.3}{\per\second}, while with the source it was \SI{36.6}{\per\second}. 
Fig.\,\ref{fig:SumOfWaveFormsMeasurements} shows a comparison of the averaged R1924A PMT waveforms, summed over the events of a typical run for the measurements taken with the $^{90}$Sr source, between the PS and the PMMA slides, as well as between the PS-WLS-2 and the PMMA-WLS-2 slides, the latter of which looks the same as for the single coated slides.

\begin{figure}[!htbp]
    \centering
    \begin{subfigure}{0.49\textwidth}
    \includegraphics[width=0.9\linewidth]{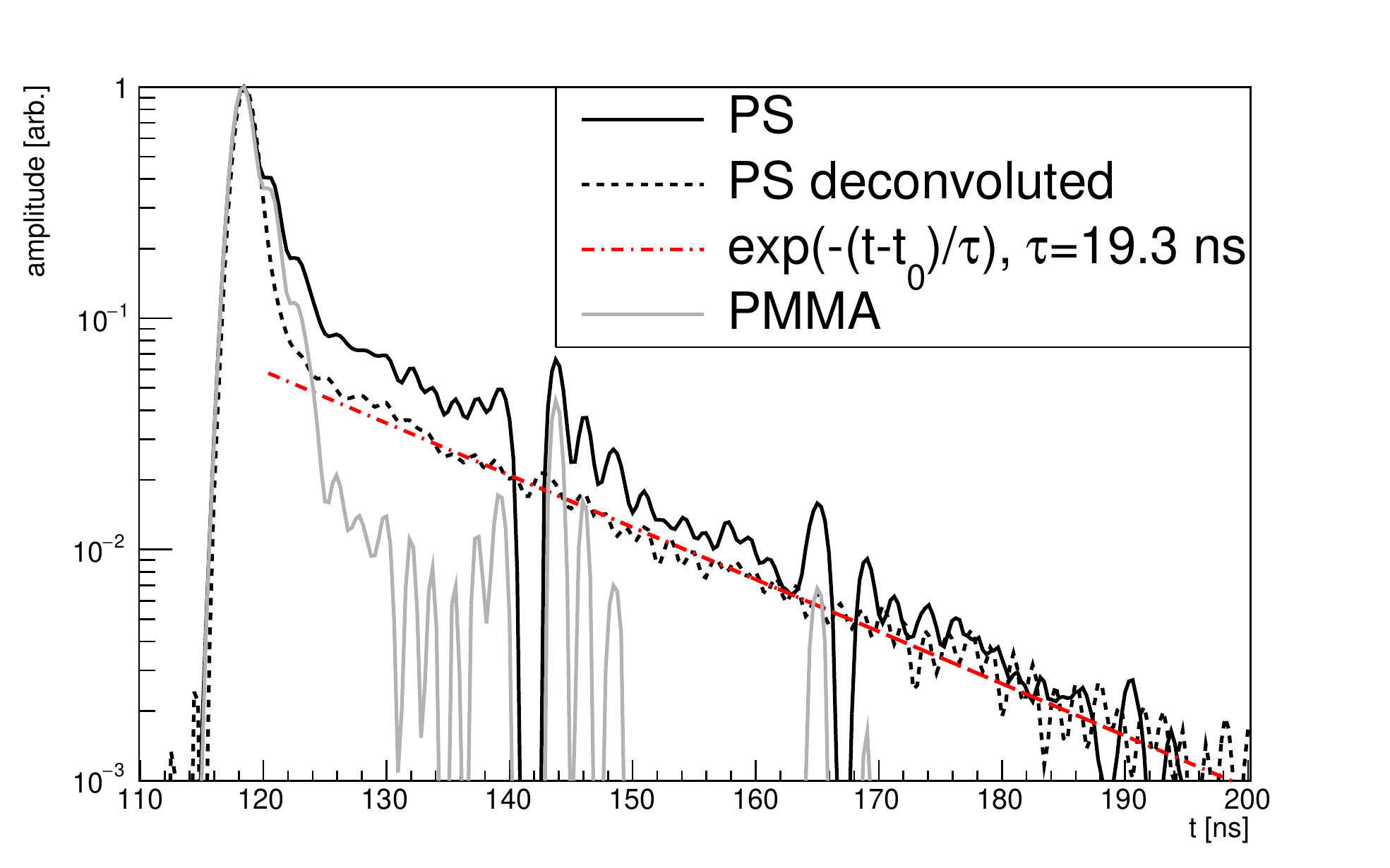} 
    \subcaption{}
    \end{subfigure}
    ~
    \begin{subfigure}{0.49\textwidth}
    \includegraphics[width=0.9\linewidth]{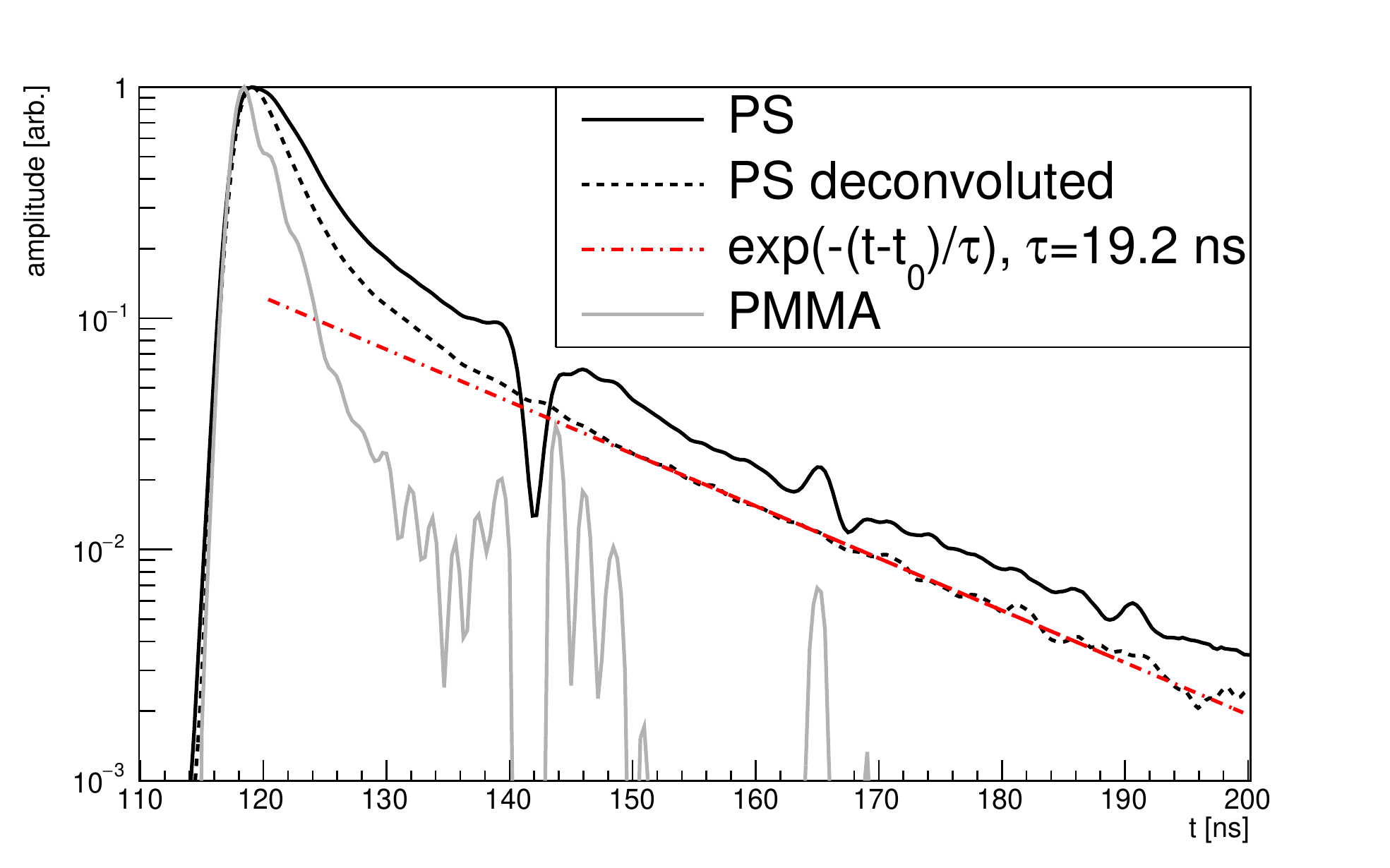} 
    \subcaption{}
    \end{subfigure}

    \caption{Averaged waveforms (same as Fig.~\,\ref{fig:SumOfWaveFormsMeasurements}) for the PS (black lines) and PMMA (grey lines) slides and the deconvolved PS waveforms (see text, dashed lines) with a fit (dash-dotted lines) of the exponential decay of the tail. (a) shows the uncoated slides and (b) the double-coated slides. The fluorescence decay time constant extracted from the fit is given in the legend.
    } 
    \label{fig:DecayTimeFit}
\end{figure}

The averaged waveforms allow us to conclude on the main light generation mechanism in the PMMA and PS slides, since Cherenkov light is emitted promptly along the particle trajectory, while scintillation light is emitted with a delay governed by the fluorescence decay time. For the uncoated slides, one observes a longer tail for PS than for PMMA (see Fig.\,\ref{fig:SumOfWaveFormsMeasurements}), which is expected for PS since not only Cherenkov photons but also scintillation photons are produced in PS. For the coated PMMA and the coated PS slides, the longer tail is even more pronounced for the PS material (see Fig.\,\ref{fig:SumOfWaveFormsMeasurements}). For pure PS,
typical fluorescence decay times in the range of \SIrange{10}{20}{\nano\second} are
reported\,\cite{Basile:1961}. This is consistent with the decay times of \SIrange{19.2}{19.3}{\nano\second} observed in the coated and uncoated PS slides, determined by fitting an exponential to the falling edge\,\cite{Basile:1961,ghiggino:1978time,condirston:1979fluorescence} of the deconvolved averaged waveforms from about \SI{20}{\nano\second} for the uncoated slide and from \SI{30}{\nano\second} for the double coated slide until \SI{70}{\nano\second} after the maximum of the waveform, as shown in Fig.\,\ref{fig:DecayTimeFit}. The fit function is exp$(-(t-t_0)/\tau)$ with the decay time $\tau$ and an offset $t_0$. The deconvolution has been performed using the response of the PMT and the read-out to a $\text{FWHM}=\SI{54}{\pico\second}$ short laser pulse (see Fig.\,\ref{fig:SumOfWaveFormsMeasurements}) as the transfer function of the setup. We can assume the photons of the short laser pulses arrive simultaneously at the PMT and the visible structure of the resulting pulse, shown in Fig.\,\ref{fig:SumOfWaveFormsMeasurements}, is the transfer function of the system including ringing and reflections. The averaged waveforms are deconvolved with this transfer function and the result is smoothed with a Gaussian with a standard deviation of \SI{800}{\pico\second} (see Appendix B of \cite{scharf:2018rad}). It should be noted that the fluorescence decay time of the WLS dye is of the order of \SI{1.5}{\nano\second} \cite{Kuzniak:2020oka} and is difficult to measure with this setup.

\begin{figure}[!htbp]
    \centering
    \subfloat[PMMA]{
        \includegraphics[width=0.45\textwidth]{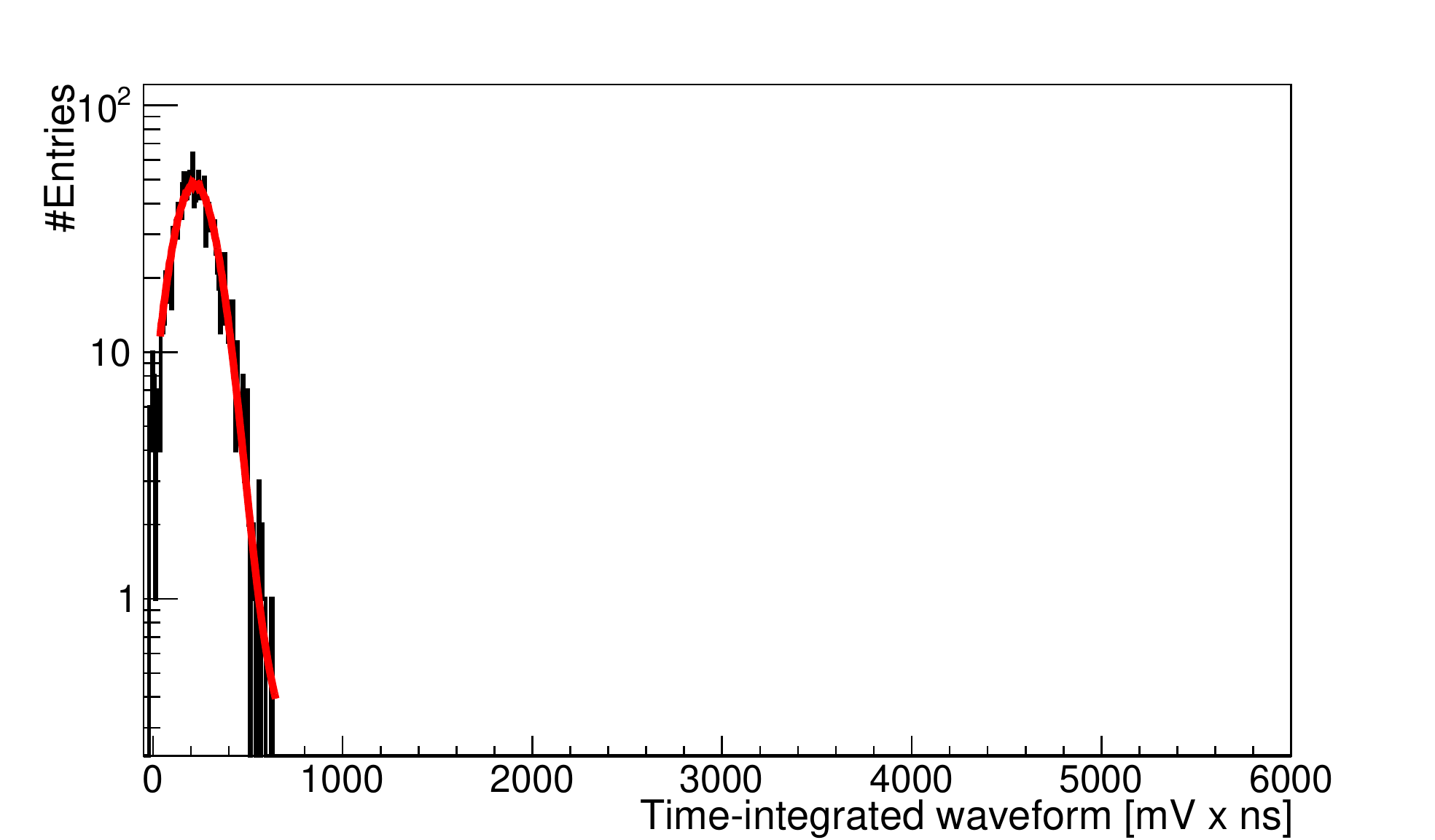}
        }
    ~ 
    \subfloat[PS]{
        \includegraphics[width=0.45\textwidth]{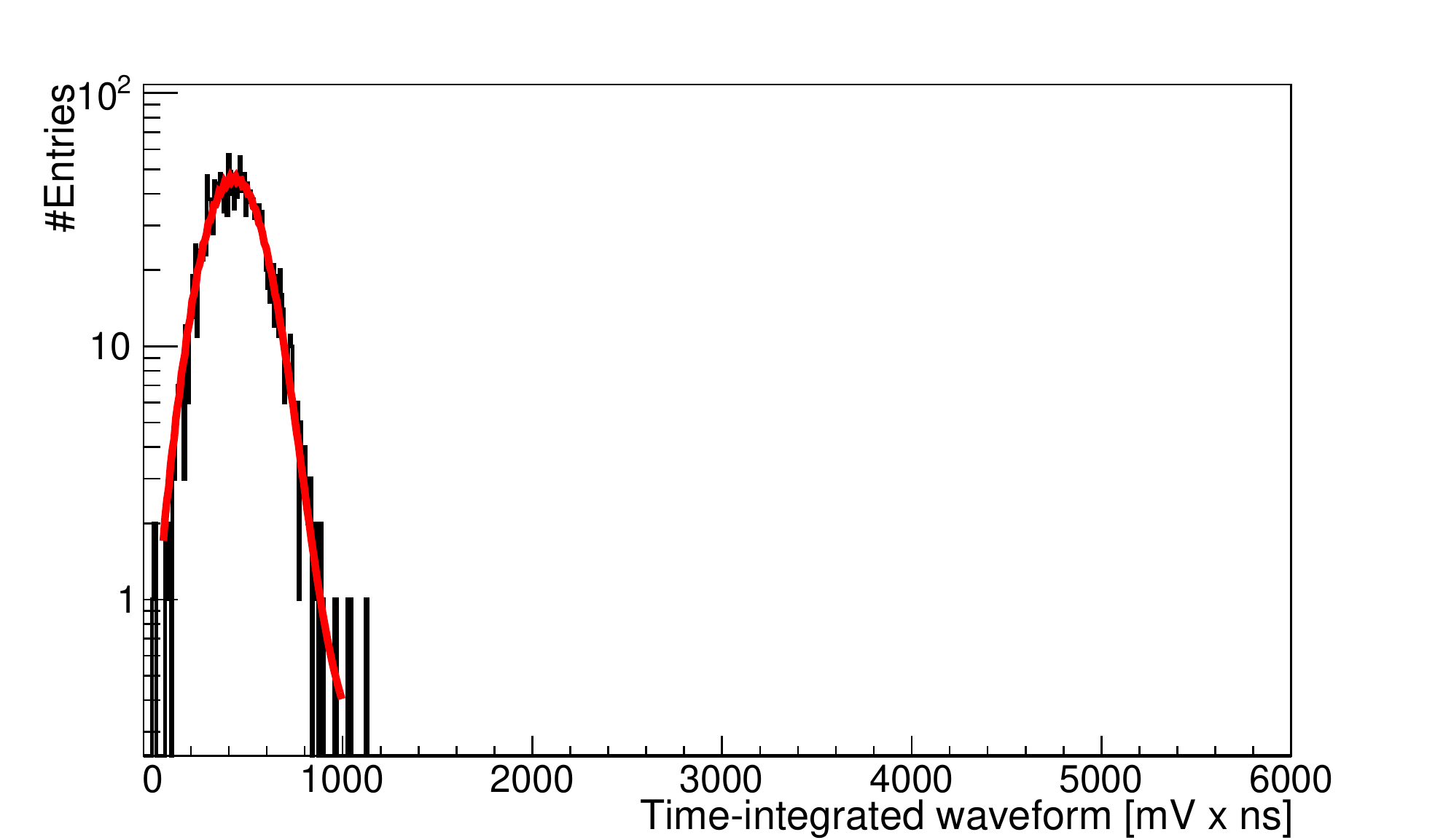}
    }
    \\[\smallskipamount]
    \subfloat[PMMA-WLS]{
        \includegraphics[width=0.45\textwidth]{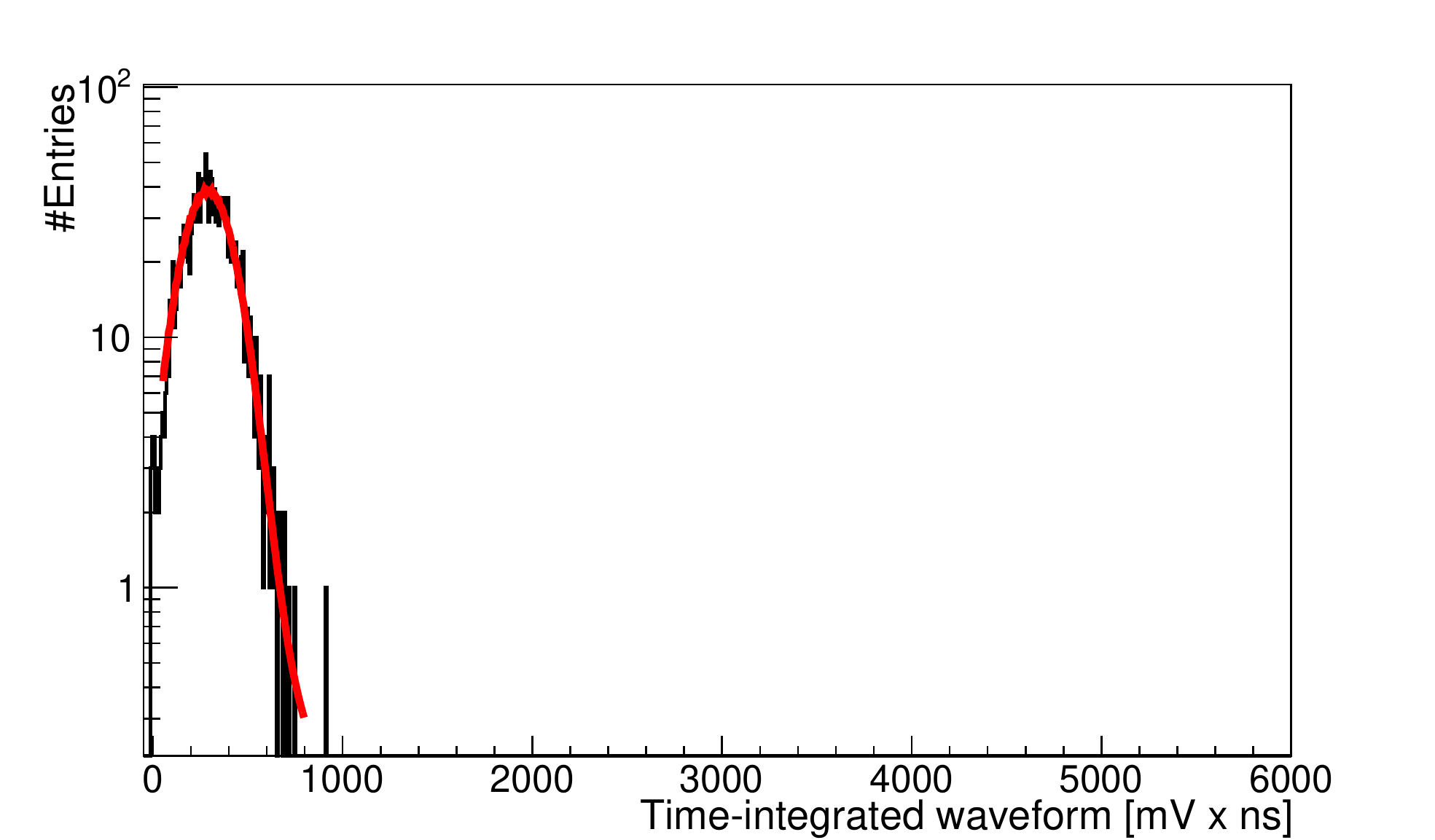}
        }
    ~
    \subfloat[PS-WLS]{
        \includegraphics[width=0.45\textwidth]{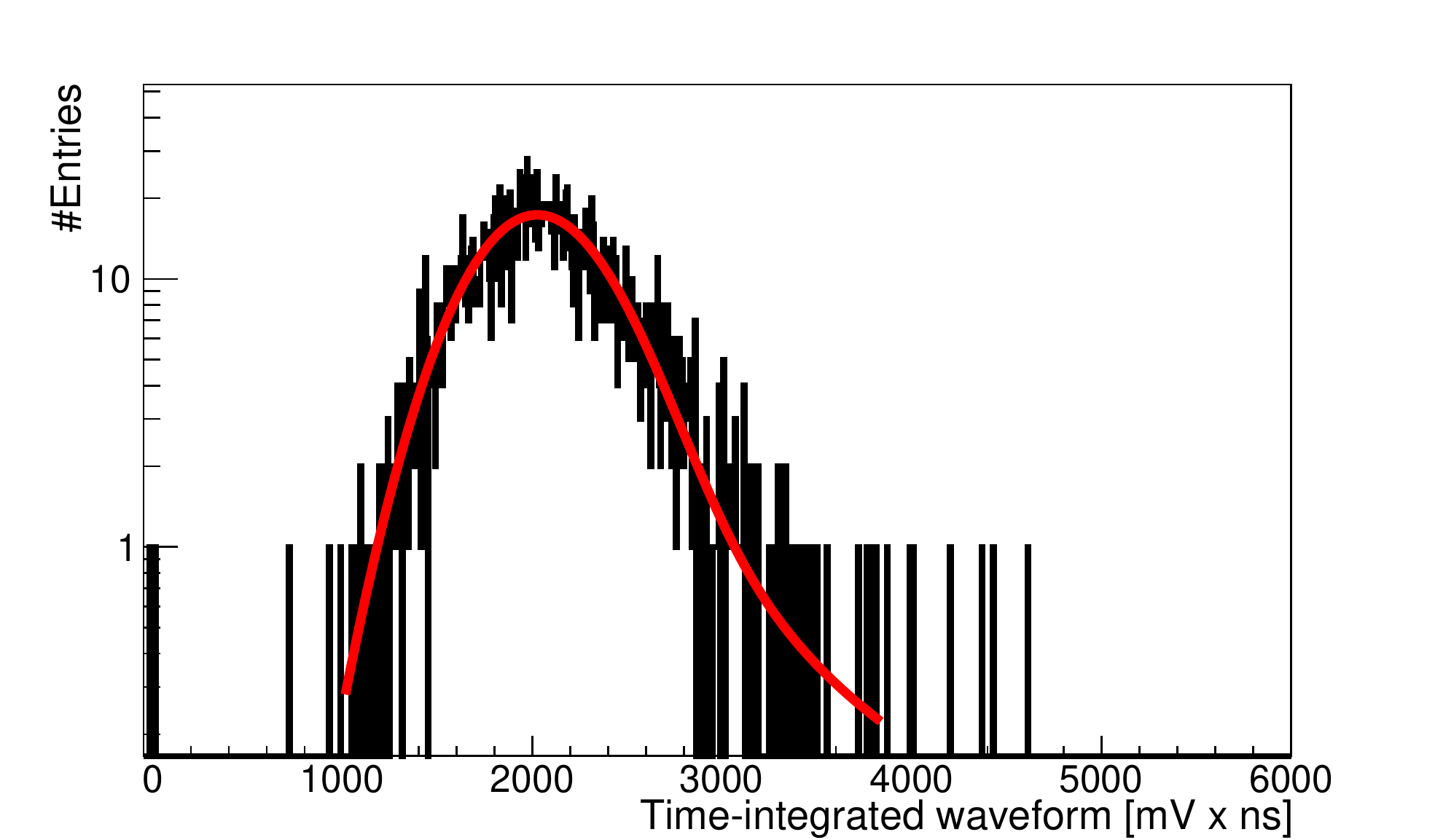}
    }
    \\[\smallskipamount]
    \subfloat[PMMA-WLS-2]{
        \includegraphics[width=0.45\textwidth]{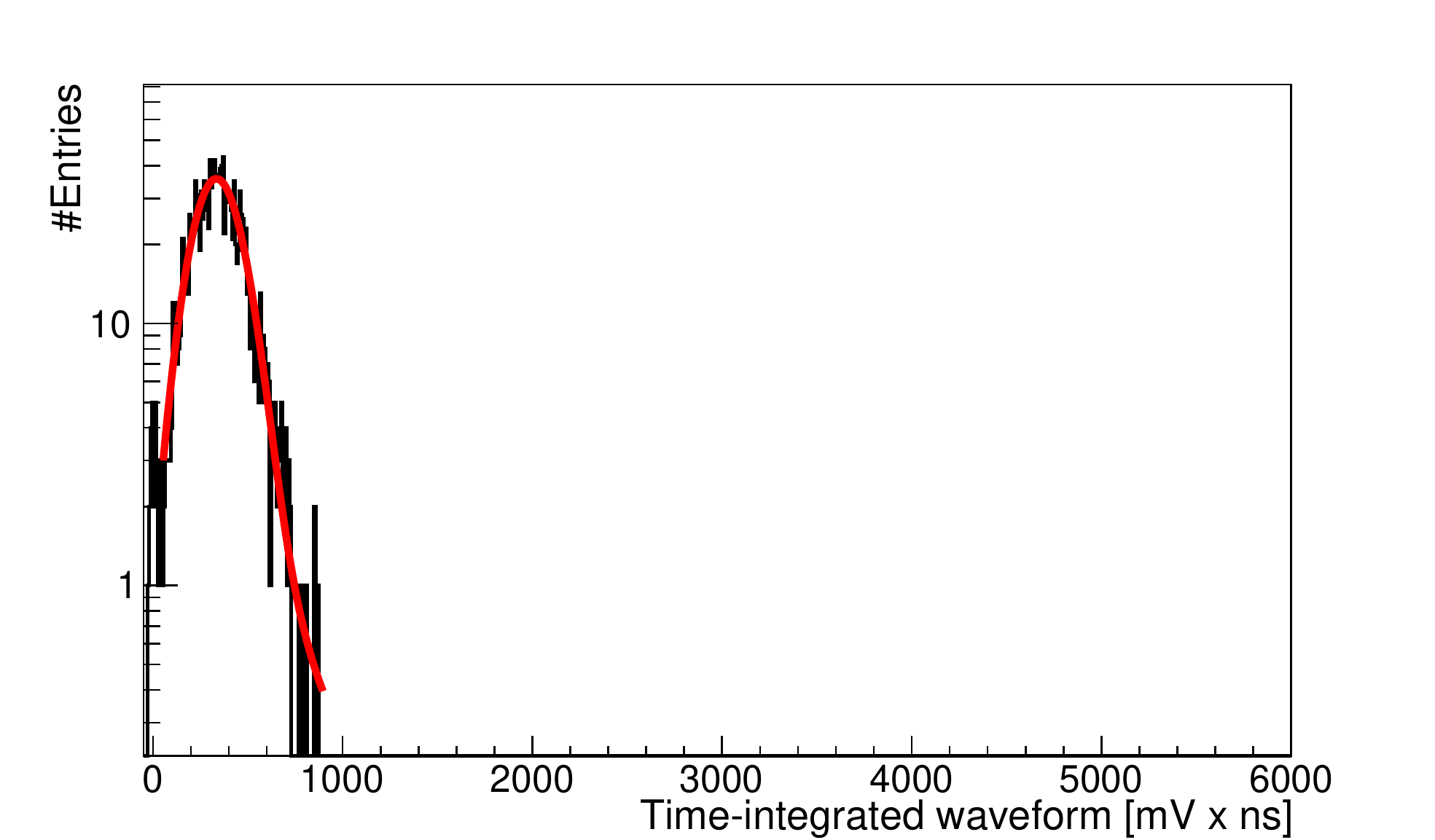}
        }
    ~
    \subfloat[PS-WLS-2 ]{
        \includegraphics[width=0.45\textwidth]{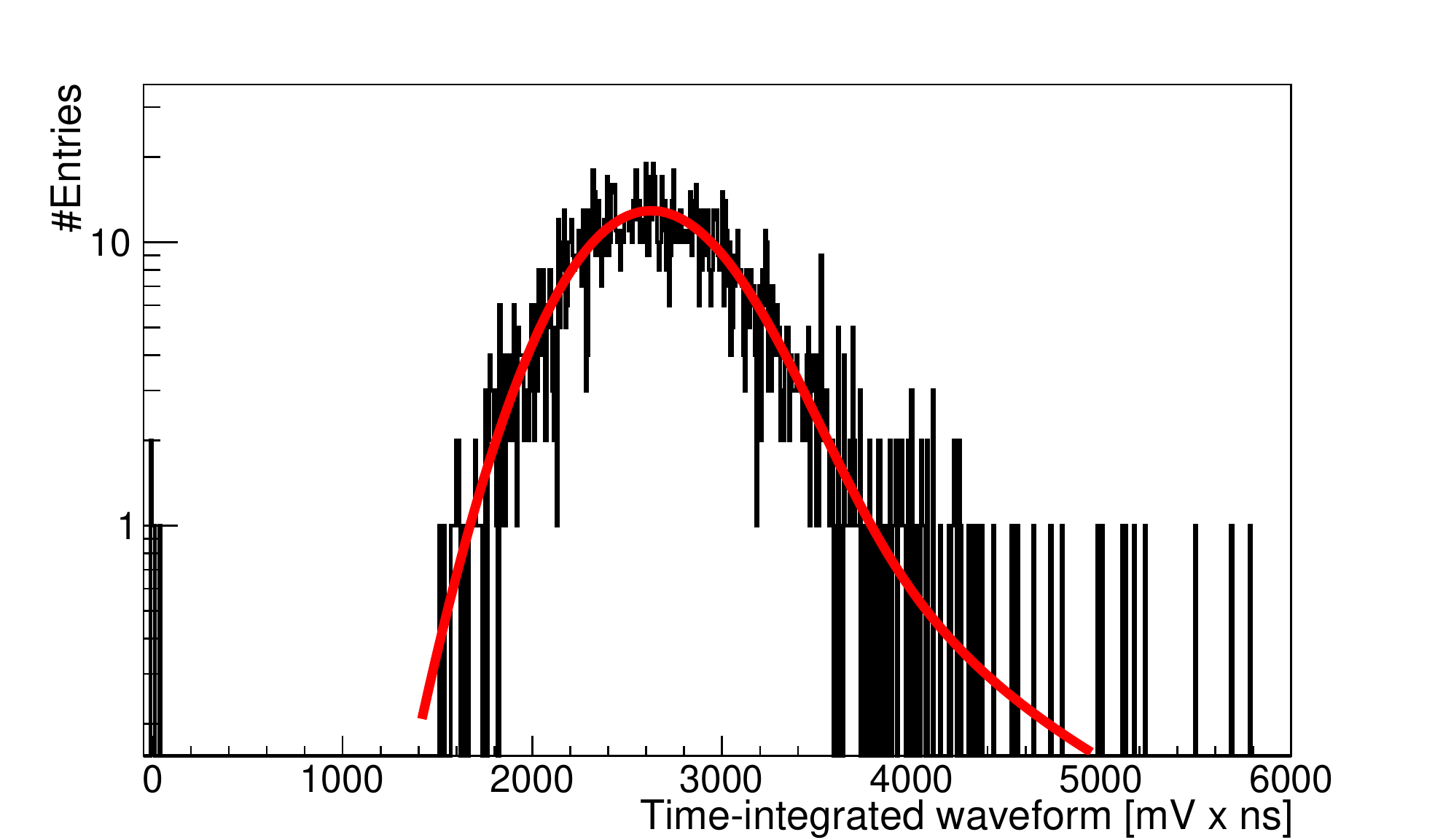}
    }
        \caption{Spectra of the time-integrated waveforms for the six plastic slides under study for measurements with the  $^{90}$Sr $\beta^{-}$ source 
        corresponding to the data shown in Fig. \ref{fig:SumOfWaveFormsMeasurements}. The curve shows the result of a fit to the measured spectrum using as a fit function the convolution of a Landau distribution with a Gaussian function
        in order to determine
        the most-probable value of the distribution, measured with the Hamamatsu R1924A PMT at a bias voltage of 1000 V.
    } 
    \label{fig:TimeIntegratedSpectraMeasurements}
\end{figure}

To quantify the number of detected photoelectrons, the waveform of each individual event was time-integrated within an integration window between \SI{110}{\nano\second} and \SI{200}{\nano\second}. 
Fig.\,\ref{fig:TimeIntegratedSpectraMeasurements} shows the time-integrated spectra obtained from the individual waveforms for the same data runs as shown in Fig. \ref{fig:SumOfWaveFormsMeasurements}.
In order to estimate the reproducibility of the optical coupling of the PMT to the plastic slide under study, two measurements were taken, the second one after decoupling and coupling again the slide.

\begin{figure}[]  
    \centering  \includegraphics[width=0.45\textwidth]{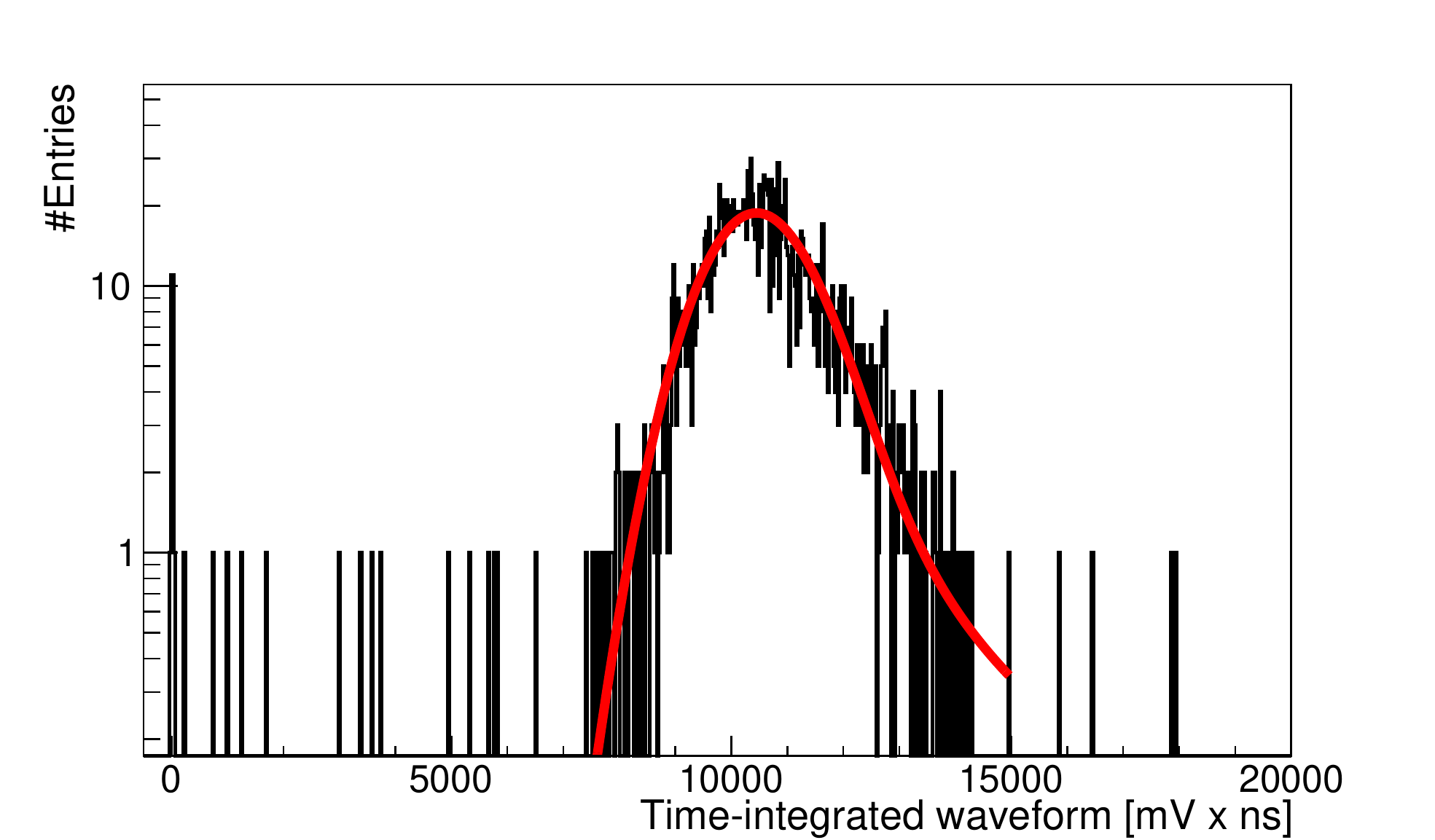}
    \caption{Spectrum of the time-integrated waveforms for the UNIPLAST slide for a measurement with the $^{90}$Sr $\beta^{-}$ source. The curve shows the result of a fit to the measured spectrum like in Fig. \ref{fig:TimeIntegratedSpectraMeasurements}, measured with the Hamamatsu R1924A PMT at a bias voltage of 900 V.
    } 
    \label{fig:TimeIntegratedSpectraMeasurementsUNIPLAST}
\end{figure}

\section{Results on the detected PE yield}
\label{Sec:Results}

Since the energy loss of a charged particle inside a thin layer of material follows a Landau distribution and the non-negligible electronic noise follows a Gaussian distribution, a convolution of a Landau distribution and Gaussian function was used to fit the time-integrated spectra of the measurements with the plastic slides (as shown in Fig. \ref{fig:TimeIntegratedSpectraMeasurements}).
To quantify the number of PE detected by the PMT, we choose the most probable value (MPV) of this fit, translating it using the equivalence between PE and mV·ns obtained from the R1924A PMT calibration measurements as described above.

Table \ref{Tab:Results} presents the PE yield obtained from the MPVs detected for the seven plastic slides under study. The PE values shown in the table correspond to the weighted mean of the MPVs of two measurements performed for each slide, divided by the calibration value. The first quoted uncertainty ($\sigma_{\scalebox{0.5}{MPV}}$) is the uncertainty of the weighted mean, quantifying the statistical uncertainty in the MPV determination. The second quoted uncertainty ($\sigma{\scalebox{0.5}{sys}}$) is calculated as the weighted standard deviation of the two measurements from the MPVs mean, allowing us to estimate whether the decoupling and coupling again the slide to the PMT introduces a significant systematic uncertainty. 
In addition, there is about a 1\% uncertainty for all measurements related to the SPE calibration value.

\begin{table}[!h]
    \centering
    \caption{PE yield obtained from the MPVs detected by the PMT for the seven plastic slides under study, using a \textsuperscript{90}Sr \mbox{$\beta$\hspace{0.5mm}\textsuperscript{-}} source. The number in brackets quotes the thickness of the plastic slide. The quoted uncertainties as described in the text correspond to the statistical uncertainty in the MPV determination ($\sigma_{\scalebox{0.5}{MPV}}$) and to the systematic uncertainty due to the optical coupling ($\sigma_{\scalebox{0.5}{sys}}$), respectively. In addition, there is about a 1\% uncertainty related to the calibration value.}
    \vspace{1.2mm}
    \begin{tabular}{m{4.5cm} P{3.2cm}}
        \hline\hline
                    & \multicolumn{1}{c}{PE $\pm \; \sigma_{\scalebox{0.5}{MPV}} \pm \sigma_{sys}$}            \\  \hline
        PMMA (\SI{5.0}{\milli\meter})       & $4.38 \pm 0.04 \pm 0.01$          \\  
        PMMA-WLS (\SI{5.00}{\milli\meter})   & $5.96 \pm 0.03 \pm 0.02$          \\  
        PMMA-WLS-2 (\SI{5.0}{\milli\meter}) & $6.92 \pm 0.04 \pm 0.04$         \\  
        \hline
        PS (\SI{4.5}{\milli\meter})         & $8.66 \pm 0.06 \pm 0.28$         \\  
        PS-WLS (\SI{4.8}{\milli\meter})     & $42.10 \pm 0.27 \pm 0.23$         \\  
        PS-WLS-2 (\SI{4.7}{\milli\meter})   & $54.58 \pm 0.28 \pm 0.34$         \\  
        \hline
        UNIPLAST (\SI{5.1}{\milli\meter})   & $506.22 \pm 1.66 \pm 7.17$         \\
        \hline\hline
    \end{tabular}
    \label{Tab:Results}
\end{table} 

Let us first discuss the results for the PMMA slide. While PMMA does not scintillate, it can result into radiation of Cherenkov light when being traversed by a charged particle with a velocity above the speed-of-light inside PMMA, $c_{PMMA}=1/n_{PMMA}$, with $n_{PMMA}=1.520$ at a wavelength of \SI{400}{\nano\meter} \cite{refr_index}. 

The number of Cherenkov photons that can be detected by the PMT depends on the incident angle of the charged particles, since the Cherenkov radiation is emitted under a defined angle with respect to the particle‘s flight direction.
Both, the Cherenkov angle and the critical angle of total reflection are
defined by the refraction index. The measurement for PMMA-WLS (PMMA-WLS-2) shows a 1.4 (1.6) times higher PE yield compared to the one for the uncoated PMMA. In absolute terms, the light yield even for the PMMA-WLS-2 slide with 6.92 PE on average is considered to be of minor interest for a charged-particle detector applications. 
A possible increase in light-output might be achieved for WLS-coated PMMA in case of UV-transparent PMMA plate material, which however is typically significantly more expensive than UV-non-transparent PMMA plate material.

The light yield detected for the uncoated PS slide (8.66 PE on average) is already two times higher than for the uncoated PMMA slide, profiting from the additional scintillation photons in PS and the higher PS transparency in the UV range. The measurement for PS-WLS (PS-WLS-2), shows a 4.9 (6.3) times higher PE yield compared to the one for the uncoated PS slide as result of the absorption of the scintillation as well as the Cherenkov UV photons in the WLS layer after a short travelling distance and the subsequent re-emission of photons in the visible range profiting from the high PS transparency in this wavelength range. In absolute terms, the detected light yield of about 55 photoelectrons in case of the double-coated PS is sufficiently large to further study and develop the concept as a detector technology, 
even if the light yield is a factor of 9.3 lower than for the UNIPLAST extruded scintillator that was used for comparison.

\section{Summary and Outlook}
\label{Sec:Summary}
This work is based on the research and development work with wavelength-shifting optical modules (WOMs), which can e.g. be used as photon detectors for large-area liquid-scintillator detectors. In this work, we modify the WOM concept in two ways: 1) Instead of a tube geometry, we consider a planar geometry. 
2) The WOM base material serves not only as a light guide but in addition as an active detector material, hence UV photons are produced either by Cherenkov radiation or scintillation inside the material. For this WOM material, we considered in particular pure polystyrene (PS), because it is intrinsically scintillating in the UV with sizeable transparency in this wavelength range and commercially available at relatively low cost.
While one can not achieve the same high light yields as for standard PS-based plastic scintillators 
(e.g. the extruded plastic scintillator material produced by UNIPLAST that was used for comparison), 
which are doped with a primary fluorophore and WLS molecules, the advantage of the material and procedure presented here is that one can produce without large efforts planar scintillators by simply dip-coating commercially available polystyrene plate material that can be cut into the desired slide geometries.

We studied the light yield of a square-shaped commercial polystyrene slide of 50 mm side length and $O(5)$ mm thickness coated with wavelength-shifter molecules
(p-terphenyl and bis-MSB), coupled to a photomultiplier, using $\beta^{-}$ particles from a $^{90}$Sr source.
Comparison measurements were made with an uncoated polystyrene slide as well as with uncoated and coated PMMA slides, the latter two of which can only produce Cherenkov light when being traversed by charged particles. The light yield produced by minimum ionizing particles detected with a Hamamatsu R1924A PMT for the single (double) coated polystyrene slide was about a factor of 4.9 (6.3) higher than the light yield detected for the uncoated polystyrene slide. The absolute yield measured with the single (double) coated polystyrene slide was about 42 (55) photoelectrons. With the same setup and using a very similar geometry, the light yield obtained with an extruded polystyrene-based plastic scintillator doped with PTP of 1.5\% concentration and POPOP of 0.01\% concentration produced by UNIPLAST was measured to be about 506 photoelectrons.

While the light yields for the coated pure polystyrene slides were much smaller than for the doped extruded plastic scintillator, they 
are still sufficiently large to make this material interesting for application in particle
detection.

On the basis of these promising results, future studies will focus on one hand on better understanding of the relative contribution of Cherenkov light and scintillation light in the used polystyrene material, and on the other hand on testing longer polystyrene strips, typically used for building larger-scale detectors, in order to quantify the effective attenuation length, the energy resolution, as well as the time and the spatial resolution of such a detector, complemented by radiation-hardness studies. The attenuation length of pure PS material, as the one under study, will likely constrain the length of such strips to O(1 m) or below.

\section{Acknowledgements}{*}
We acknowledge the fruitful discussion with all the individuals from JGU Mainz and DESY involved in establishing the WOM as a sensor for IceCube.
We thank Emil List-Kratochvil and Andreas Opitz from the Institute of Physics at Humboldt University of Berlin for helpful discussions on increasing the WLS layer thickness in the WLS dip-coating process and for providing access to the transmission spectrophotometer PerkinElmer Lambda 950. We thank Axel Hagedorn from DESY for providing Tyvek paper.
We acknowledge the support from BMBF via the High-D consortium.

\end{document}